\documentclass[a4paper,11pt]{article}
\pdfoutput=1 
\usepackage{jcappub}
\usepackage{graphicx}
\pdfoutput=1 
\usepackage{color}
\usepackage[dvipsnames]{xcolor}

\title{Physical effects involved in the measurements of neutrino masses with future cosmological data}

\author[1]{Maria Archidiacono,}
\author[1]{Thejs Brinckmann,}
\author[1]{Julien Lesgourgues}
\author[1,2]{ and Vivian Poulin}

\affiliation[1]{Institute for Theoretical Particle Physics and Cosmology (TTK), \\ RWTH Aachen University, D-52056 Aachen, Germany.}
\affiliation[2]{LAPTh, Universit\'e Savoie Mont Blanc \& CNRS, BP 110,\\ F-74941 Annecy-le-Vieux Cedex, France.}

\abstract{Future Cosmic Microwave Background experiments together with upcoming galaxy and 21-cm surveys will provide extremely accurate measurements of different cosmological observables located at different epochs of the cosmic history. The new data will be able to constrain the neutrino mass sum with the best precision ever. In order to exploit the complementarity of the different redshift probes, a deep understanding of the physical effects driving the impact of massive neutrinos on CMB and large scale structures is required. The goal of this work is to describe these effects, assuming a summed neutrino mass close to its minimum allowed value. We find that parameter degeneracies can be removed by appropriate combinations, leading to robust and model independent constraints. A joint forecast of the sensitivity of Euclid and DESI surveys together with a CORE-like CMB experiment leads to a $1\sigma$ uncertainty of $14$~meV on the summed neutrino mass. However this particular combination gives rise to a peculiar degeneracy between $M_\nu$ and the optical depth at reionization. Independent constraints from 21-cm surveys can break this degeneracy and decrease the $1\sigma$ uncertainty down to $12$~meV.}

\begin{document}

\hfill{\small TTK-16-44, LAPTH-062/16}

\maketitle

\section{Introduction}

A wide program of future cosmological experiments is planned or proposed, in order not only to pin down cosmological parameters, but also to shed light on fundamental physics related to cosmology. These cosmological experiments include high precision galaxy redshift surveys, such as Euclid, DESI, WFIRST (see \cite{Font-Ribera:2013rwa} and references therein), high precision cosmic shear surveys, such as Euclid and LSST, and finally  Cosmic Microwave Background experiments aimed at more accurate polarization measurements, such as CORE~\cite{Bouchet:2011ck,DeZotti:2016qfg,core-proposal} and CMB-Stage IV~\cite{Allison:2015qca,Hlozek:2016lzm}. 

However, besides experimental sensitivity, parameter constraints are limited by degeneracies: a degeneracy indicates the ability of one parameter to mimic the effect of another parameter on a particular observable, making it impossible to disentangle them and to corner the value of each parameter separately. The key approach to tackle this problem consists in a joint analysis of complementary probes with different degeneracy directions in parameter space. For that reason, the next step in the era of precision cosmology will be based on the synergy of high- and low- redshift probes.

One of the parameters that will benefit from such an approach is the neutrino mass sum (hereafter $M_\nu$). Indeed the impact of massive neutrinos on cosmological observables comes from a very special effect: light massive neutrinos behave as radiation before their non-relativistic transition, while afterwards they gradually become a matter component; therefore their impact on cosmological probes at different redshifts is closely related to their mass.


The neutrino mass effects have been widely studied in the literature~\cite{Bashinsky:2003tk,Hannestad:2010kz,Lesgourgues:2012uu,Lesgourgues:1519137} and their impact on CMB and large scale structures on linear scales is well known. Even on non-linear scale, the neutrino mass effect is better understood thanks to recent progress in N-body simulations~\cite{Brandbyge:2008rv,Bird:2011rb,Brandbyge:2010ge,AliHaimoud:2012vj}.

However, neutrino cosmology is about to face a revolution for two reasons. 

First of all, current upper bounds on the neutrino mass sum are getting closer and closer to the minimum value allowed by the inverted hierarchy $M_\nu \sim 0.11$~eV~\cite{Palanque-Delabrouille:2015pga,Cuesta:2015iho,Aghanim:2016yuo}.
Thus, future experiments will look at ultra-light neutrinos that became non-relativistic in a relatively recent cosmological epoch.
Very small neutrino masses will have a different effect on the cosmological evolution and, thus, a different impact on cosmological observables.
For instance, even if the majority of neutrinos with $m_\nu \leq 600$~meV go non-relativistic after photon decourpling, a small number of them (contributing to the low momentum tail of the phase space distribution) are already partially non-relativistic at decoupling. For $m_\nu \sim 300$~meV this could still have small effects, but not for $m_\nu \sim 60$~meV. Similarily, neutrinos becoming relativistic soon after photon decoupling can produce a distortion of the CMB temperature spectrum through the early Integrated Sachs-Wolfe effect, but again this can only be significant for masses of a few hundreds of meV. Instead, the neutrinos studied in this paper could have individual masses of at most 100~meV.

Secondly, future galaxy surveys will reach a very high sensitivity on very small scales. As for now, the use of small scale data is limited by the uncertainty on non linear structure formation, which is difficult to model, especially in presence of massive neutrinos~\cite{Archidiacono:2015ota,Carbone:2016nzj,Castorina:2015bma,Dupuy:2015ega,Fuhrer:2014zka}. A major theoretical goal in the next few years will be to provide a better understanding of the processes governing clustering on small scales. Having the non linear effects under control, we will be able to exploit small scale data in order to break degeneracies. The neutrino mass effects are already important on linear scales, but by including smaller and smaller scales one would have a better lever arm and improve the constraints on $M_\nu$.


The aim of this work is to investigate the physical effects induced by massive neutrinos as they will be unveiled by future cosmological data.

We will pay specific attention to the correlation between $M_\nu$ and other cosmological parameters, and show that directions of degeneracy are very sensitive to probes of the cosmic history at different epochs. For some combinations of CMB and Large Scale Structure data sets, a correlation between $M_\nu$ and $\tau_\mathrm{reio}$ has already been observed in references \cite{Liu:2015txa,Allison:2015qca}, but its interpretation is far from obvious and requires a detailed investigation. This correlation is very important, for the reason that independent measurements of the optical depth by 21cm surveys will lead to a remarkable improvement on the sensitivity to the neutrino mass~\cite{Liu:2015txa}. We will confirm this expectation with a dedicated forecast showing that even the minimum allowed value of the summed neutrino mass could be detected at the 5$\sigma$ level in a time scale of about ten years.

The paper is organized as follows: In section~\ref{sec:cmb_intro}, \ref{sec:bao} and \ref{sec:lss} we study in detail the effect of a variation of the summed neutrino mass in CMB, Baryonic Acoustic Oscillation (BAO) and Large Scale Structure (LSS) observables, respectively. In particular we will carefully describe and explain the degeneracies with other relevant cosmological parameters. In section~\ref{sec:results} we will present the results of our Markov Chain Monte Carlo forecast of the sensitivity of future CMB, BAO, LSS and 21cm experiments. Finally in section~\ref{sec:conclusions} we will draw our conclusions.
\section{Effect of a small neutrino mass on the CMB}
\label{sec:cmb_intro}
\subsection{General parameter degeneracies for CMB data}

In the minimal, flat, 6-parameter $\Lambda$CDM model, it is well-known that the CMB temperature and polarisation unlensed spectra are determined by a number of effects\footnote{For a review of these effects, see e.g. \cite{Hu:1995em}, section 5.1 of \cite{Lesgourgues:1519137}, and \cite{Howlett:2012mh}.}, which remain identical as long as one fixes quantities usually depending on distance and density ratios, such as:
\begin{itemize}
\item the sound horizon angular scale $\theta_s(z_\mathrm{dec})=\frac{d_s(z_\mathrm{dec})}{d_A(z_\mathrm{dec})}$ at decoupling, 
\item the diffusion angular scale $\theta_d(z_\mathrm{dec})=\frac{d_d(z_\mathrm{dec})}{d_A(z_\mathrm{dec})}$ at decoupling, 
\item the baryon-to-photon ratio $R_\mathrm{dec}=\left. \frac{3 \rho_b}{4 \rho_\gamma} \right|_\mathrm{dec}$ at decoupling,
\item the redshift of radiation-to-matter equality $z_\mathrm{eq}=\frac{\rho_m^0}{\rho_r^0}-1$,
\item the redshift of  matter-to-cosmological-constant equality $z_\mathrm{\Lambda}=\left(\frac{\rho_\Lambda^0}{\rho_r^0}\right)^{1/3}-1$. 
\end{itemize}
The CMB spectra also depend on a few extra parameters, like the scalar amplitude and tilt $(A_s, n_s)$ and the optical depth at reionization $\tau_\mathrm{reio}$. However, bearing in mind that the small-$\ell$ (large angular) part of the spectra is loosely constrained due to cosmic variance, the parameters $z_\Lambda$, $A_s$ and $\tau_\mathrm{reio}$ are always less constrained by CMB data than $(\theta_s, \theta_d, R, z_\mathrm{eq}, n_s)$, and also than the combination $A_s e^{-2 \tau_\mathrm{reio}}$ giving the overall spectrum normalisation on small angular scales. The fact that we actually measure lensed CMB spectra gives extra information on the amplitude and slope of the matter power spectrum $P(k,z)$ at low redshift: in practice, this increases the sensitivity to the parameters $(A_s, z_\Lambda)$, which enter into the normalisation of $P(k,z)$. 

Adding neutrino masses into the model leads to several new effects studied extensively in the literature~\cite{Lesgourgues:1519137,Hou:2012xq,Howlett:2012mh}:
\begin{description}
\item[(a)] Neutrino masses affects the background expansion history. If we rely on standard assumptions for the photon and background densities ($T_\mathrm{cmb}=2.726$~K, $N_\mathrm{eff}=3.046$) and further fix $\omega_b$ and $\omega_\mathrm{cdm}$, the changes in the background evolution caused by neutrino masses are confined to late times. Then, the values of $d_s(z_\mathrm{dec})$, $d_d(z_\mathrm{dec})$, $R_\mathrm{dec}$ and $z_\mathrm{eq}$ are preserved, and the neutrino masses only impacts the angular diameter distance (and therefore, $\theta_s$ and $\theta_d$ in an equal way) and $z_\Lambda$ (and hence, the loosely constrained late ISW effect). It is even possible to choose an appropriate value of the cosmological constant for each set of neutrino masses, in order to keep a fixed $d_A(z_\mathrm{dec})$: in that case, the impact of neutrino masses on the background is confined to variations of $z_\Lambda$ and of the late ISW effect, and cannot be probed accurately due to cosmic variance, unless external non-CMB datasets come into play.
\item[(b)] At the perturbation level, massive neutrinos interact gravitationally with other species and produce small distortions in the CMB peaks. For individual neutrino masses $m_\nu$ smaller than $\sim 600$~meV, the neutrinos become non-relativistic after recombination: in that case the distortions can only be caused by the early ISW effect, and affect the CMB temperature spectrum in the multipole range $50<\ell<200$ \cite{Lesgourgues:2012uu,Hou:2012xq,Lesgourgues:1519137}. Note that this neutrino-mass-induced early ISW effect takes place even if the redshift of equality is kept fixed: it is different from the redshift-of-equality-induced early ISW effect, which affects the height of the first CMB peak in the range $100 < \ell < 300$.
\item[(c)] Finally, at the lensing level, massive neutrinos slow down the growth of small-scale structure (leading to the well-known suppression factor $1-8\omega_\nu/\omega_m$ in the small-scale matter power spectrum at redshift zero) and globally decrease the impact of CMB lensing: the peaks are less smoothed and the damping tail less suppressed~\cite{Lewis:2006fu}.
\end{description}
All these effects have played a role in previous constraints on neutrino masses from CMB data alone, or combined with other probes. Interestingly, while the sensitivity of CMB instruments increases with time, different effects come to dominate the neutrino mass constraints: early ISW effects {\bf (b)} with WMAP alone~\cite{Hinshaw:2012aka}, lensing effects {\bf (c)} with Planck alone~\cite{Ade:2015xua}, and background effects {\bf (a)} when combining CMB data with direct measurements of $H_0$~\cite{Riess:2016jrr}. There are now several combinations of cosmological probes giving a 95\%CL upper bound on the summed mass $M_\nu\equiv \sum m_\nu$ of the order of 120~meV to 150~meV  \cite{Palanque-Delabrouille:2015pga,Cuesta:2015iho,Aghanim:2016yuo,Giusarma:2016phn}, while neutrino oscillation data enforces $M_\nu \geq 60$~meV  at 95\%CL \cite{Gonzalez-Garcia:2015qrr}. The remaining conservatively allowed window is so narrow, $\Delta M_\nu \sim 90$~meV, that the impact of a realistic variation of the neutrino masses on the CMB is getting really small. Our purpose in section~\ref{sec:cmb} is to study precisely this impact, and to understand the degeneracy between $M_\nu$ and other parameters when using future CMB data only, specifically for the very low mass range $60$~meV$ < M_\nu < 150$~meV. This requires some preliminary remarks in section~\ref{sec:cmb_data}.

\subsection{CMB data definition}
\label{sec:cmb_data}

The discussion of degeneracies is meaningless unless we specify which data set, and which experimental sensitivities, we are referring to.
In this paper, we take as a typical example of future CMB data a next-generation CMB satellite similar to the project COrE+, submitted to ESA for the call M4.
A new version of CORE was recently submitted again for the call M5~\cite{DeZotti:2016qfg,core-proposal}, with a small reduction of the instrumental performances, mainly in angular resolution. However, COrE+ and CORE-M5 are very similar, and the conclusions of this paper would not change significantly by adopting the CORE-M5 settings. 

When displaying binned errors in $C_l$ plots, and when doing MCMC forecasts with mock data and synthetic likelihoods, we assume that this  
CORE-like experiment is based on 9 frequency channels with sensitivity and beam angles given in footnote\footnote{
Assumed specifications for a COrE+ - like experiment: frequencies in GHz: [100, 115, 130, 145, 160, 175, 195, 220, 255];
$\theta_{\rm fwhm}$ in arcmin: [8.4, 7.3, 6.46, 5.79, 5.25, 4.8, 4.31, 3.82, 3.29];
temperature sensitivity in [$\mu$K arcmin] : [6.0, 5.0, 4.2, 3.6, 3.8, 3.8, 3.8, 5.8, 8.9];
polarisation sensitivity in [$\mu$K arcmin] : [8.5, 7.0, 5.9, 5.0, 5.4, 5.3, 5.3, 8.1, 12.6].
}. We mimic the effect of sky masking by adopting a Gaussian likelihood with an overall rescaling by a sky fraction $f_\mathrm{sky}=0.70$. 

Our dataset consists primarily of temperature and E-mode polarisation auto-correlation and cross-correlation spectra 
$C_\ell^{TT}, C_\ell^{EE}, C_\ell^{TE}$.
To get more information on CMB lensing, one can either analyse B-mode maps (in absence of significant primordial gravitational waves, the B-mode only comes from CMB lensing and foregrounds) and add the $C_\ell^{BB}$ spectrum to the list of observables; or perform lensing extraction with a quadratic or optimal estimator~\cite{Matsumura:2016sri}, and add the CMB lensing potential spectrum $C_\ell^{\phi \phi}$ to the list of observables (equivalently one could use the deflection spectrum $C_\ell^{d d} = \ell (\ell+1) C_\ell^{\phi \phi}$ ). We cannot use both $C_\ell^{BB}$ and $C_\ell^{\phi \phi}$ in the likelihood: the same information would be counted twice. Here we choose to use the lensing potential spectrum, which better separates the contribution of different scales to lensing, that would be mixed in the $C_\ell^{BB}$ spectrum by some integration kernel. To give an example, we will see in Figure~\ref{fig:difftt} (bottom plots) that the neutrino mass effect on $C_\ell^{\phi \phi}$ is more pronounced on small angular scales, while in the lensed $C_\ell^{BB}$ this effect would be nearly independent of $\ell$\footnote{However, we will also see that {\it within the range in which error bars are small}, the neutrino mass effect on $C_\ell^{\phi \phi}$ is also {\it nearly} $\ell$-independent, so we may expect that trading $C_\ell^{\phi \phi}$ against $C_\ell^{BB}$ in the likelihood would have a minor impact on our conclusions.}.

So the CMB data set that we have in mind consists in measurements for $C_\ell^{TT}, C_\ell^{EE}, C_\ell^{TE}, C_\ell^{\phi \phi}$, with a synthetic Gaussian likelihood similar to that in \cite{Perotto:2006rj}, and a  lensing extraction error spectrum $N_\ell^{\phi\phi}$ based on the quadratic estimator method~\cite{Okamoto:2003zw} for the EB estimator. In the likelihood, we keep the lensed $C_\ell^{TT}, C_\ell^{EE}, C_\ell^{TE}$. Indeed, unlike $C_\ell^{BB}$, these spectra are only weakly affected by lensing, and the lensing information redundency between lensed temperature/polarisation spectra and the  $ C_\ell^{\phi \phi}$ spectrum  is small enough for being negligible at the instrumental sensitivity level of a CORE-like experiment \cite{core-params}.

\subsection{Degeneracies between very small $M_\nu$'s and other parameters with CMB data only}
\label{sec:cmb}

\begin{figure*}[h]
\begin{tabular}{ll}
\includegraphics*[width=0.49\linewidth]{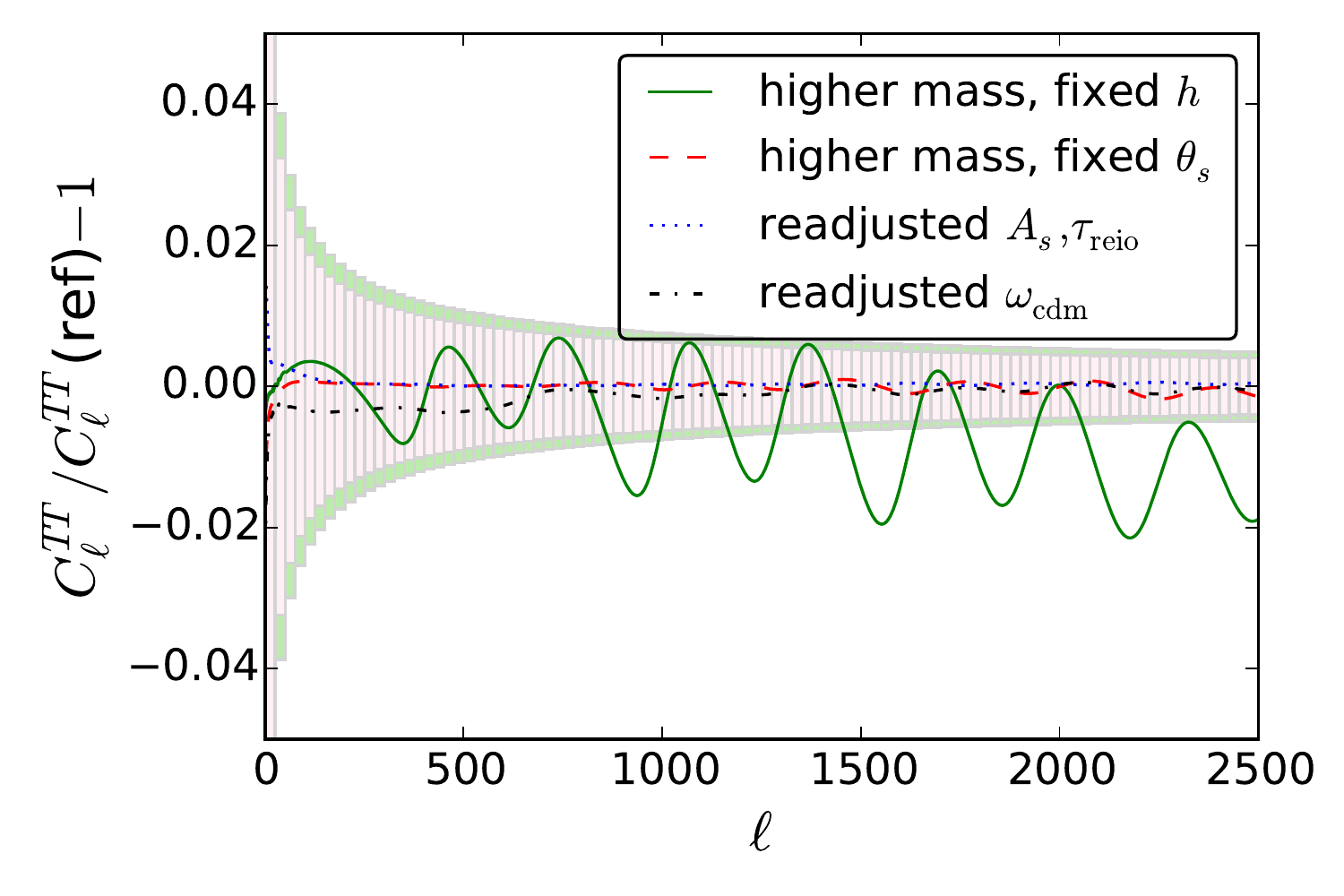}&
\includegraphics*[width=0.49\linewidth]{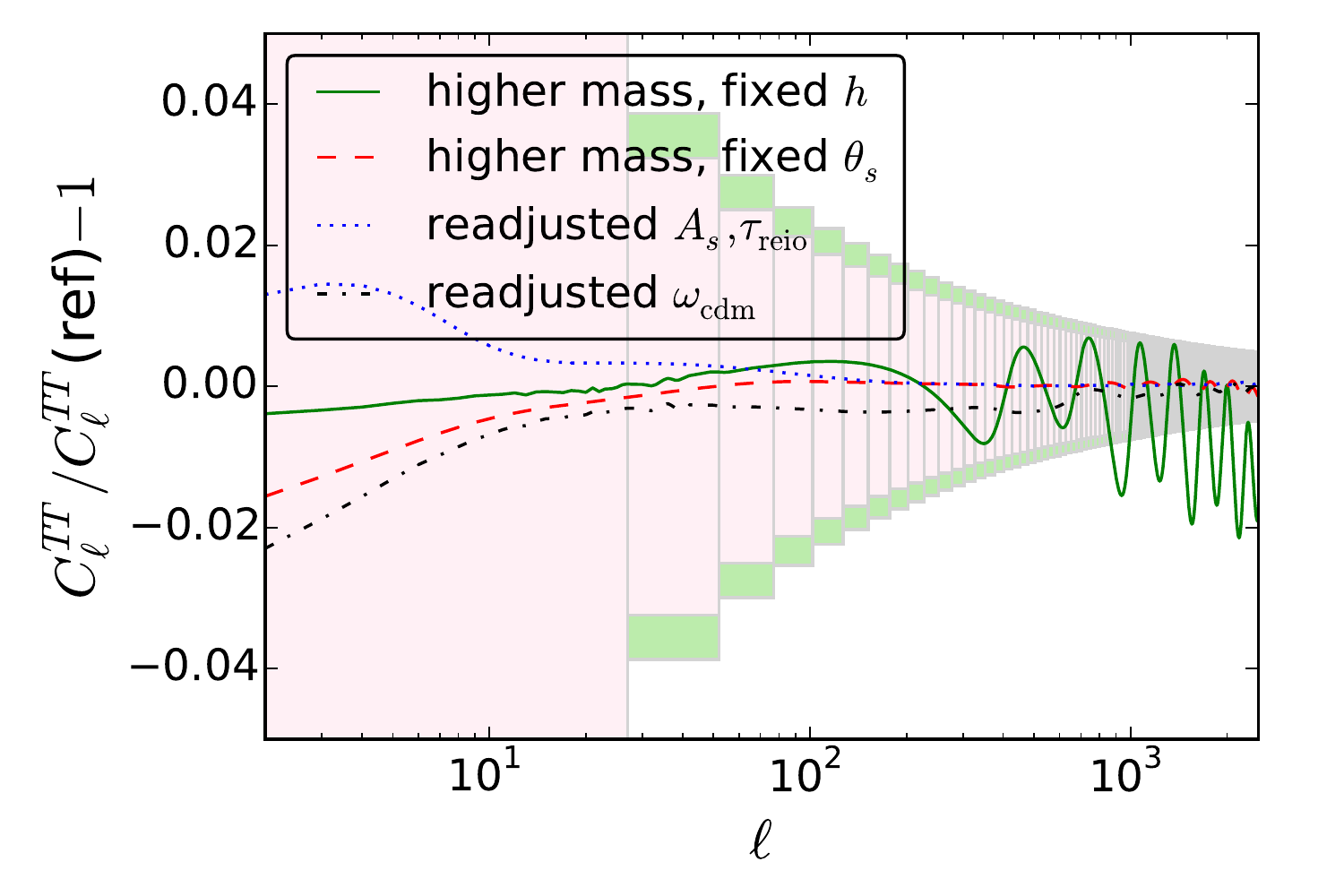}\\
\includegraphics*[width=0.49\linewidth]{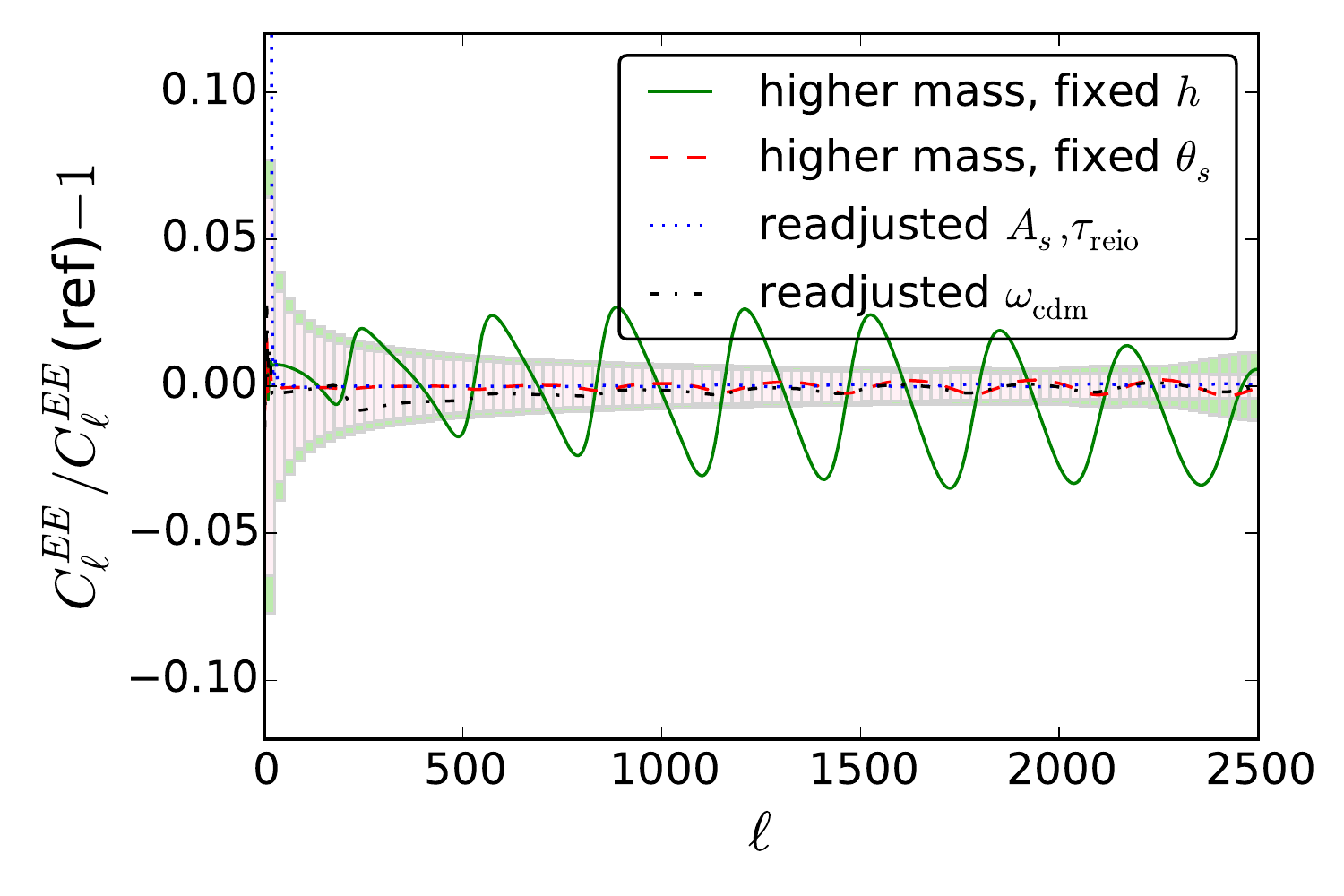}&
\includegraphics*[width=0.49\linewidth]{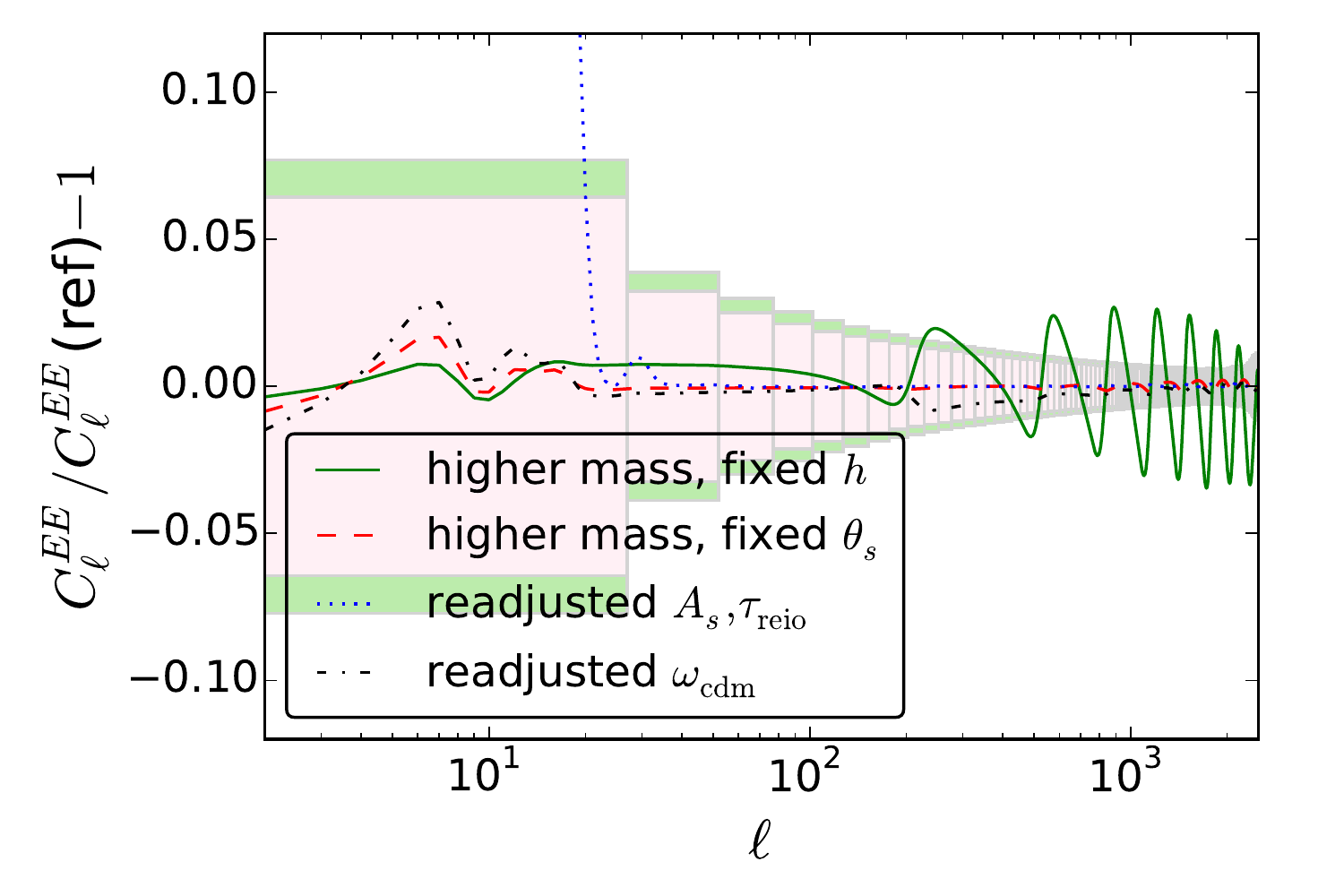}\\
\includegraphics*[width=0.49\linewidth]{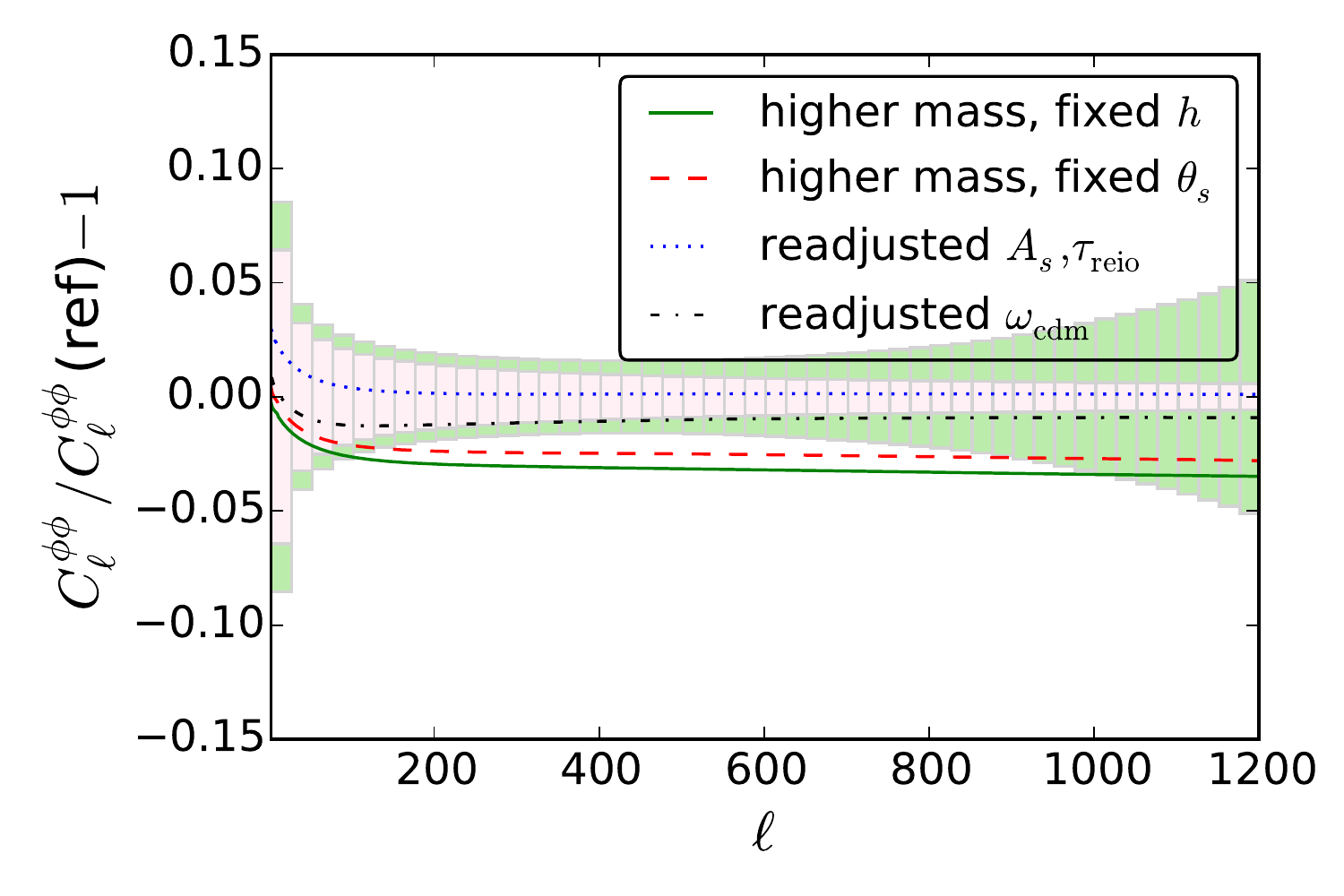}&
\includegraphics*[width=0.49\linewidth]{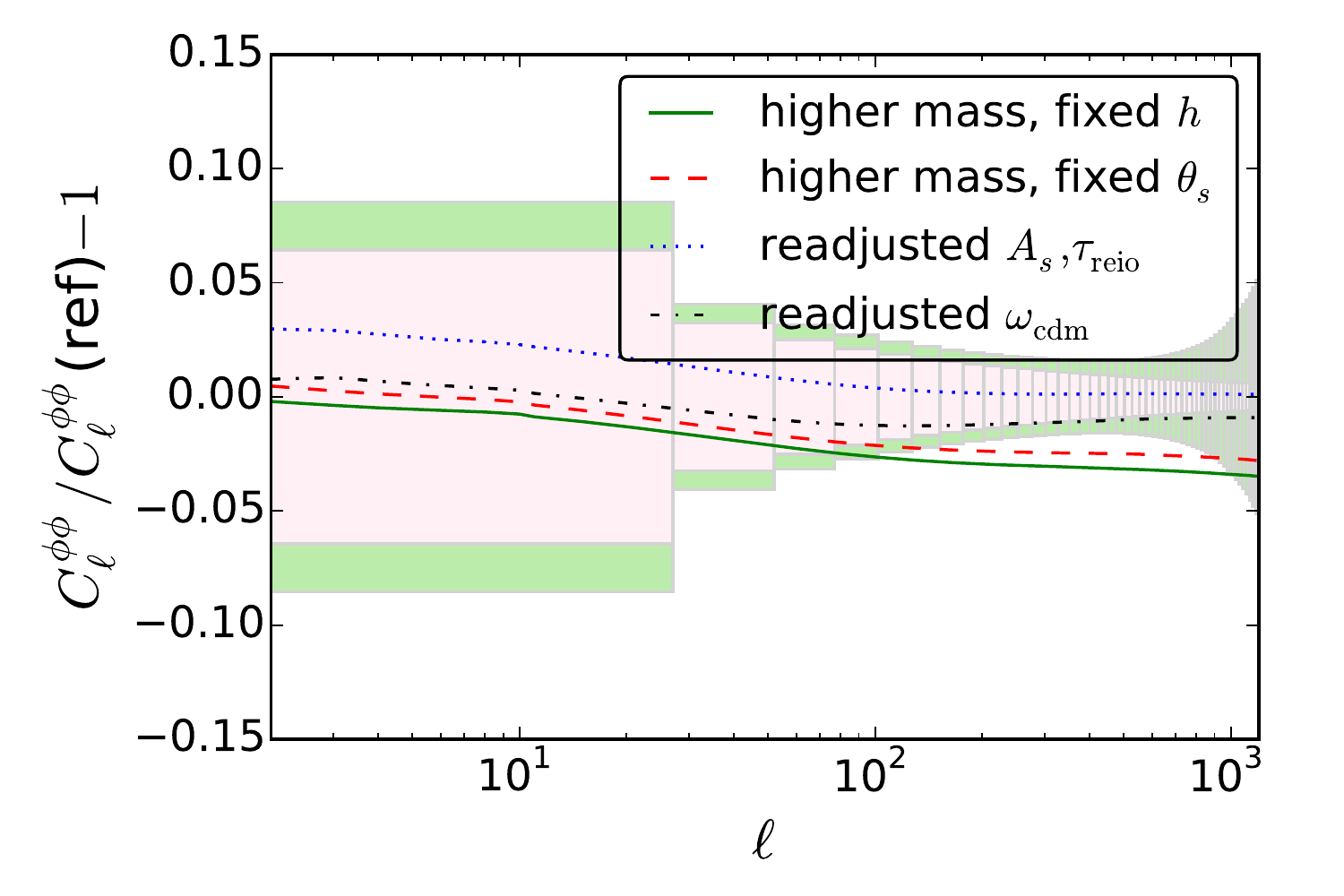}
\end{tabular}
\caption{ \label{fig:difftt}
Relative change in the CMB spectra induced by increasing the summed neutrino mass from $M_\nu=60$~meV to $M_\nu=150$~meV. The plots show the residuals of the lensed $TT$ (top), lensed $EE$ (middle) and lensing potential (bottom) power spectrum, as a function of multipoles $\ell$ with a linear (left) or logarithmic (right) scale. The light/pink and darker/green shaded rectangles refer, respectively, to the binned noise spectrum of a cosmic-variance-limited or CORE-like experiment, with linear bins of width $\Delta \ell=25$. 
The physical baryon density $\omega_b$ and the scalar spectral index $n_s$ are kept fixed. In the first case (green solid line) the value of the Hubble constant is fixed at the reference value, while in all the other cases (labeled as fixed $\theta_s$) $h$ decreases in order to keep $\theta_s$ consistent with the reference model. Moreover, in the third case (dotted blue line), we tried to compensate for the changes in the lensing spectrum by increasing $A_s$, and in the fourth case (dotted-dashed black) we aim at the same result by increasing $\omega_\mathrm{cdm}$.
}
\end{figure*}

We will discuss the impact of increasing the neutrino mass, while keeping various parameters or combination of parameters fixed.  We illustrate this discussion with the plots of Figure~\ref{fig:difftt}, showing the spectrum ratio between different models sharing a summed mass $M_\nu=150$~meV and a baseline model\footnote{
Our discussion is general and the value of cosmological parameters for the baseline model is unimportant. We choose Planck-inspired values, 
$
\left\lbrace \omega_b, \omega_{\rm cdm}, h,n_s, A_s,\tau_{\rm reio},M_\nu  \right\rbrace \, = \,
 \left\lbrace 0.02214, 0.12070, 0.6663, 0.9624, 2.12 \times 10^{-9}, 0.0581, 0.06 \, {\rm eV} \right\rbrace
$, giving an angular sound horizon at recombination $\theta_s$ (which defines the angular scale of the CMB acoustic peaks) roughly equal to $100\, \theta_s=1.04075$. In this work, our total neutrino mass $M_\nu$ is assumed to be shared equally among the three species, like in the degenerate (DEG) model. This choice is not random: it is motivated by the fact that the cosmological impact of different mass splittings is negligible, at least as long as one compares some DEG, NH (Normal Hierarchy) and IH (Inverted Hierarchy) models all sharing the same total mass $M_\nu$. This is not necessarily true when comparing them with a model with ``one massive, two massless neutrinos'', which departs by a much larger amount at the level of the matter power spectrum \cite{Lesgourgues:2012uu,Lesgourgues:1519137}. Hence, by varying the mass $M_\nu$ of the DEG model, we obtain results which apply at the same time, in very good approximation, to the two realistic cases NH and IH.}
with $M_\nu=60$~meV. The plots shows the residuals of the lensed $TT$ (top), lensed $EE$ (middle) and lensing potential (bottom) power spectrum, as a function of multipoles $\ell$ with a linear (left) or logarithmic (right) scale. The light/pink and darker/green shaded rectangles refer respectively to the binned noise spectrum of a cosmic-variance-limited or CORE-like experiment, with linear bins of width $\Delta \ell=25$. All spectra are computed with the Boltzmann solver {\sc class}\footnote{\tt http://class-code.net}~\cite{Lesgourgues:2011re,Blas:2011rf,Lesgourgues:2011rh}, version 2.5.0, with the high precision settings {\tt cl\_permille.pre}.

Our discussion will also be illustrated by the results of Monte Carlo Markov Chains (MCMC) forecasts for our CORE-like experiment: Figure~\ref{fig:cmb_2d} gives the 2D probability contours for the pairs of parameters most relevant to our discussion. The MCMC forecasts are done with the {\sc MontePython} package\footnote{\tt http://baudren.github.io/montepython.html}~\cite{Audren:2012wb}.

The main conclusions can be reached in four steps:
\begin{enumerate}
\item
We first assume that we increase neutrino masses with respect to the baseline model, while keeping the parameters $\left\lbrace \omega_b, \omega_{\rm cdm}, h,n_s, A_s,\tau_{\rm reio}\right\rbrace$ fixed (green solid curve in Figure~\ref{fig:difftt}). Given the discussion in point {\bf (a)}, we expect that this is not a very clever choice, because the angular diameter distance is not preserved. So if the baseline model is a good fit to the data, the new model will be discrepant. Indeed, by looking especially at the top left and middle left plots in Figure~\ref{fig:difftt}, we see even-spaced oscillations signaling a change in the angular diameter distance, and the residuals are far above the  instrumental noise.
\item
We then perform the same increase in $M_\nu$, but now with a fixed angular diameter distance to recombination, which means that $\left\lbrace \omega_b, \omega_{\rm cdm}\right\rbrace$ are still fixed, but $h$ varies. With {\sc class}, this is easily achieved by keeping the input parameter $100\theta_s$ constant. Since the early cosmology and the sound horizon at decoupling are fixed, fixing $\theta_s$ means adjusting $H_0$ and the angular diameter distance for each $M_\nu$. Then, the angular diffusion scale $\theta_d$ is also automatically fixed. 
In Figure~\ref{fig:difftt}, this transformation corresponds to the dashed red residuals. As expected, the previous oscillations disappear in the residuals. The only visible effects are much smaller oscillations, some tilt at large $\ell$ due to a different level of CMB lensing, and a tilt at small $\ell$ due to a different late ISW effect. However, both effects are below cosmic variance. We conclude that, the measurement of the temperature and E-mode spectra alone does not allow us to distinguish between $M_\nu=60$~meV and 150~meV, and that in  a CMB analysis the parameters $(M_\nu, H_0)$ are inevitably correlated, as it is well known, and illustrated by the upper left plot in Figure~\ref{fig:cmb_2d}.
\begin{figure*}[h]
\begin{tabular}{ll}
\includegraphics*[width=0.49\linewidth]{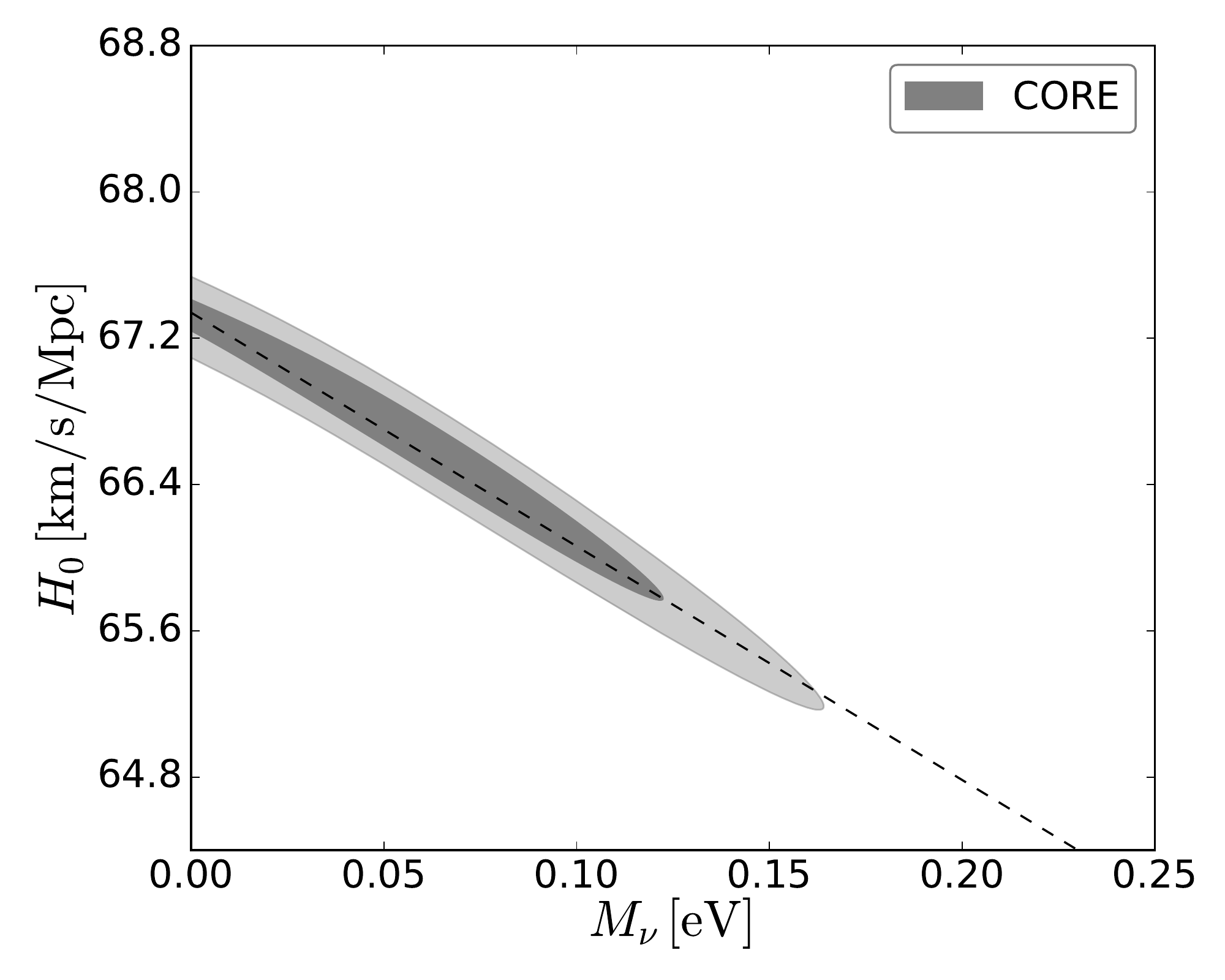}&
\includegraphics*[width=0.49\linewidth]{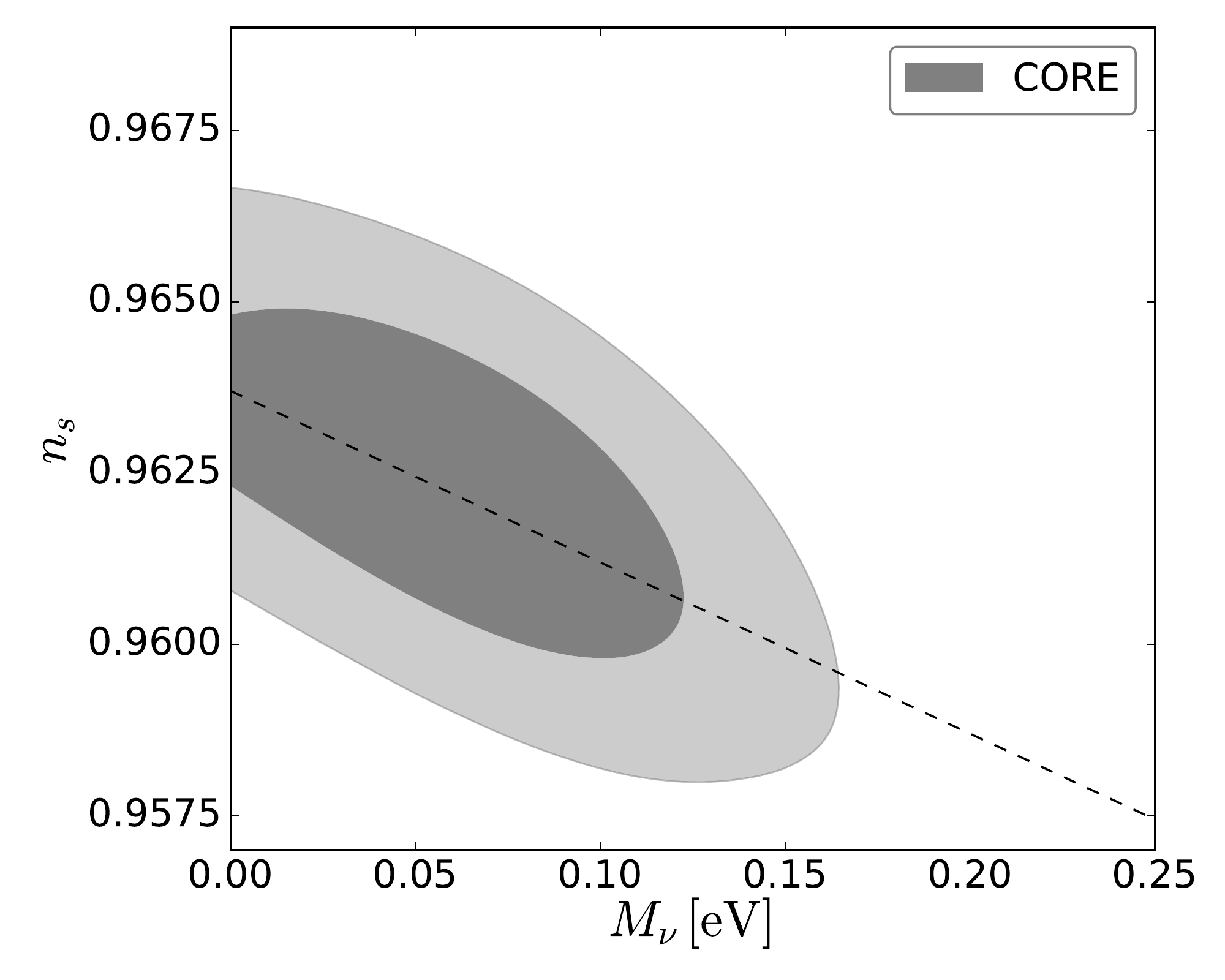}\\
\includegraphics*[width=0.49\linewidth]{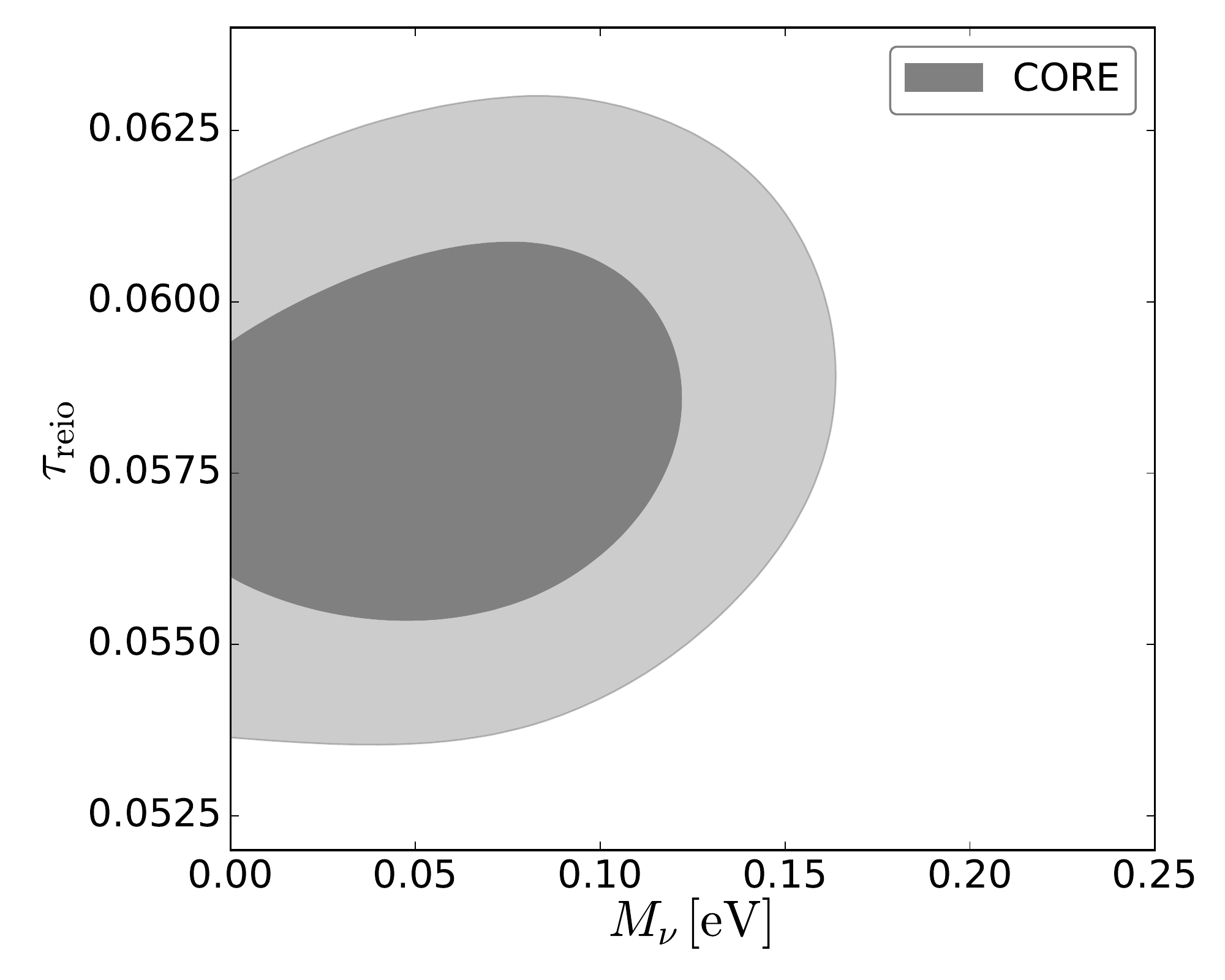}&
\includegraphics*[width=0.49\linewidth]{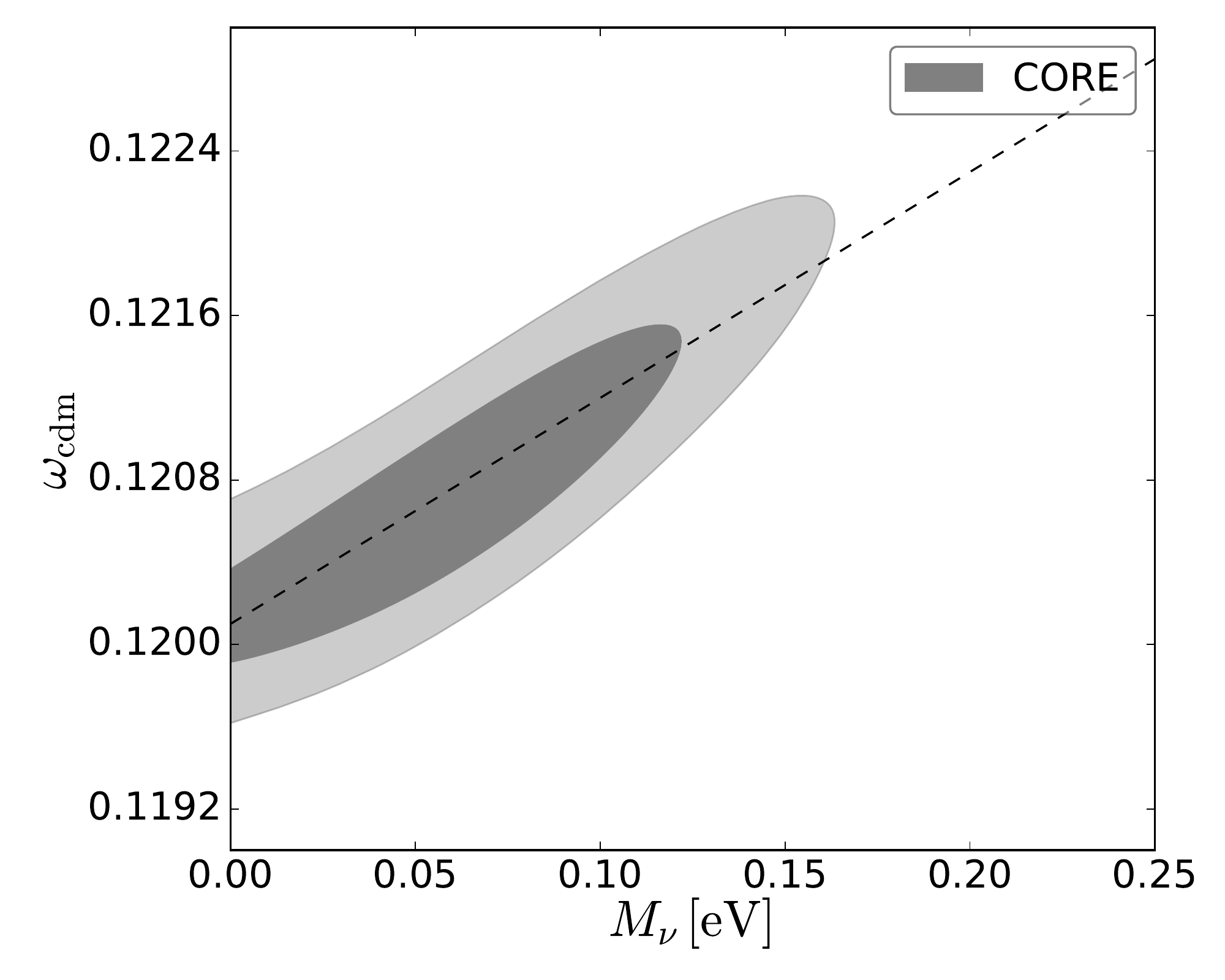}
\end{tabular}
\caption{ \label{fig:cmb_2d}
68\% and 95\% CL posterior probability contour levels for different pairs of parameters, for an MCMC forecast of the sensitivity of a CORE-like experiment to the parameters of a 7-parameter model ($\Lambda$CDM plus total neutrino mass $M_\nu$). The CMB data is assumed to consist of measurements of the TT, EE, TE and lensing potential spectra. 
}
\end{figure*}

We can try to quantify this correlation. A simple numerical exercise shows that in order to keep the same value of $\theta_s$ while fixing  $\left\lbrace \omega_b, \omega_{\rm cdm}\right\rbrace$ and varying $M_\nu$, one finds a correlation
\begin{equation}
\Delta h \simeq -0.09 \, \left( \frac{\Delta M_\nu}{1~\mathrm{eV}} \right)~.
\label{cmb:eq:mnu-h1}
\end{equation}
We will come back to this relation later, and show that the correlation angle changes slightly when other effects are taken into account.

We now look at the bottom plots in Figure~\ref{fig:difftt}, showing variations in the lensing potential spectrum. The dashed red line is consistent with the fact that a higher neutrino mass implies more suppression in the small-scale matter power spectrum $P(k,z)$, and hence in the large-$\ell$ lensing potential spectrum $C_l^{\phi \phi}$. A comparison with the instrumental errors show that this effect is potentially relevant: the dashed red residual is outside the 68\% error bars in about ~30 consecutive bins, leading to a $\chi^2$ increase by many units. It is also visible that the neutrino mass effect would be detectable only in a range given roughly by 
$50 \leq \ell \leq 800$, in which the effect is nearly equivalent to a suppression by some $\ell$-independent factor (by about 3\% in our example).

Hence, we see that a CORE-like CMB experiment could in principle discriminate between $M_\nu=60$~meV and 150~meV, and that the effect of the neutrino mass with fixed $\left\lbrace \omega_b, \omega_{\rm cdm}, \theta_s, n_s, A_s,\tau_{\rm reio}\right\rbrace$ can be simply summarised as an apparent mismatch between the normalisation of the TT,TE,EE spectra and that of the CMB lensing spectrum. To check whether the distinction can be made in reality, and not just in principle, we must think whether the variation of other cosmological parameters could cancel this effect, and lead to new parameter correlations with $M_\nu$.

As explained in references~\cite{Pan:2014xua,Ade:2015zua}, in a pure $\Lambda$CDM model with no massive neutrinos, the dependence of the global amplitude of $C_\ell^{\phi\phi}$  on the cosmological parameters is given approximately by:
\begin{equation*}
\ell^4C_\ell^{\phi\phi} \propto A_s \left( \Omega_m^{0.6}h\right)^{2.5} \;\,\,(\ell > 200),
\end{equation*}
and in terms of $\omega_m$
\begin{equation*}
\ell^4C_\ell^{\phi\phi} \propto A_s  \omega_m^{3/2} h^{-1/2} \;\,\,(\ell > 200),
\end{equation*}
plus an additional minor dependence on $\Omega_m$.
If we include massive neutrinos, the linear growth of structure becomes scale dependent, thus the exact impact of $M_\nu$ on $C_\ell^{\phi\phi}$ 
is $\ell$-dependent, but only by a small amount in the range constrained by observations. Anyway, given that the neutrino mass slows down the growth of cold dark matter perturbations, we can generally assume:
\begin{equation}
\ell^4C_\ell^{\phi\phi} \propto A_s \omega_m^{3/2} h^{-1/2} M_\nu^{-\alpha},
\label{eq:clff}
\end{equation}
with $\alpha > 0$. This qualitative result shows that in order to compensate an increase of $M_\nu$, we have a priori two possibilities: increasing $A_s$, or increasing $\omega_m$. We will explore them one after each other in the next points, and arrive at interesting conclusions. 


\item We have the possibility to increase $A_s$ in order to compensate for the neutrino mass effect in $C_\ell^{\phi\phi}$, while keeping $A_s e^{-2 \tau_{\rm reio}}$ fixed, in order to have the same overall normalisation of the large-$\ell$ temperature and polarisation spectra. Hence, this transformation implies a higher reionisation optical depth $ \tau_{\rm reio}$. We could expect that, this change in the optical depth is unobservable due to cosmic variance, which would mean that there is a parameter degeneracy at the level of CMB data, and that the three parameters $(M_\nu, A_s, \tau_{\rm reio})$ are correlated.

This turns out not to be the case. In our example, the higher neutrino mass shifts the lensing potential down by 3\%. This could be compensated by increasing $A_s$ by 3\% as well, and shifting $\tau_{\rm reio}$ by $\Delta \tau_{\rm reio} = \frac{1}{2} \log 1.03 \simeq 0.015$. This is a very big shift compared to the expected sensitivity of a CORE-like experiment, $\sigma(\tau_{\rm reio}) \simeq 0.002$. Hence this degeneracy should not be present.

This is illustrated by the third set of curves (dotted blue) in Figure~\ref{fig:difftt}. We estimated numerically the reduction factor for $C_{400}^{\phi \phi}$ in the second model (red dashed). We increased $A_s$ by exactly this factor, keeping $A_s e^{-2 \tau_{\rm reio}}$ fixed. The new model has a much larger reionisation bump in $C_\ell^{EE}$, with a residual largely exceeding the error bars.

The lower left plot in Figure~\ref{fig:cmb_2d} brings the final confirmation that in a global fit of CMB data, with lensing extraction included, there is no significant correlation between $M_\nu$ and $\tau_{\rm reio}$.

At this point, we still expect that very small neutrino masses could be accurately measured by CMB data alone, unless the other way to compensate for the neutrino mass effect in the lensing potential (by increasing $\omega_m$) works better  than increasing $A_s$, and does lead to some parameter degeneracy. This is what we will explore in the final step of this discussion.


\item Considering that $\omega_b$ is accurately determined by the first peak ratios, we can increase $\omega_{\rm m}= \omega_{\rm b} + \omega_{\rm cdm}$ by enhancing $\omega_{\rm cdm}$ only. It is difficult to infer analytically from equation (\ref{eq:clff}) the amount by which $\omega_{\rm cdm}$ should be enhanced in order to cancel the effect of $M_\nu$ in the lensing potential, because during the transformation, we must keep $\theta_s$ fixed; since $\theta_s$ depends on both $h$ and $\omega_{\rm m}$, the Hubble parameter will also change.
In the example displayed in figure~\ref{fig:difftt}, we found numerically the factor by which we should increase $\omega_{\rm cdm}$ (with fixed $\omega_{\rm b}$ and $\theta_s$), in order to nearly cancel the neutrino mass effect in the lensing power spectrum. This leads to the dotted-dashed black curve. In the lensing potential plots (bottom), the new residual is back inside the cosmic variance band.

The problem with the previous attempt was that changing $\tau_{\rm reio}$ had ``side effects'' (namely, on the reionisation bump) potentially excluded by the data. Increasing $\omega_{\rm cdm}$ also has ``side effects'': it affects the redshift of radiation/matter equality $z_{\rm eq}$, and hence the amplitude of the first two peaks (through gravity boost effects and through the early ISW effect); it also affects the redshift of matter/$\Lambda$ equality $z_\Lambda$ and the late ISW effect; and finally, it has a small impact on the angular diameter distance. All these effects can be identified by looking at the details of the dotted-dashed black residuals in figure~\ref{fig:difftt}. The key point is that a tiny enhancement of $\omega_{\rm cdm}$ is enough to compensate for the neutrino mass effect in $C_{\ell}^{\phi \phi}$, in such way that the ``side effects'' all remain well below cosmic variance.  Hence, we expect a parameter degeneracy between $M_\nu$ and $\omega_{\rm cdm}$ when using CMB data alone, that will compromise the accuracy with which the neutrino mass can be pinned down, and lead to a correlation
between these parameters. We notice that this correlation between $M_\nu$ and $\omega_\mathrm{cdm}$ is completely driven by CMB lensing. Removing lensing extraction would diminish the correlation factor. The residual correlation would be due to the lensing of the $C_\ell^{TT}$ spectrum (related to the tiny deviation of the black dot dashed line from the red dashed line on small scales in the top left panel of figure~\ref{fig:difftt}), and it would disappear with delensing.

This is confirmed by the lower right plot in Figure~\ref{fig:cmb_2d}: in a global fit of CMB data, we obtain a degeneracy direction approximately parametrised by the slope of the dashed curve in that plot,
\begin{equation}
\Delta \omega_{\rm cdm} = 0.01 \, \Delta M_\nu \sim \Delta \omega_\nu~.
\label{cmb:eq:mnu-cdm}
\end{equation}
Which is exactly the relation we used in figure~\ref{fig:difftt}, when transforming to the fourth model (dotted-dashed black curves).
\end{enumerate} 


We can reach the main conclusion of this section: for CMB data alone (including lensing extraction), there is no significant parameter degeneracy between $(M_\nu, A_s, \tau_{\rm reio})$, but there is one between $M_\nu$ and $\omega_{\rm cdm}$. This is the most pronounced parameter degeneracy involving the neutrino mass
when the cosmological model is parametrised by $\left\lbrace \omega_b, \omega_{\rm cdm}, \theta_s,n_s, A_s,\tau_{\rm reio}\right\rbrace$, and the correlation is given approximately by equation~(\ref{cmb:eq:mnu-cdm}).

If instead the model is parametrised by 
$\left\lbrace \omega_b, \omega_{\rm cdm}, h,n_s, A_s,\tau_{\rm reio}\right\rbrace$, for the obvious reasons discussed previously, there is an additional clear correlation between $M_\nu$ and $h$. We return to the correlation factor, that we estimated before to be given by equation~(\ref{cmb:eq:mnu-h1}). This equation is actually not a very good fit of the contours in the upper left plot of Figure~\ref{fig:cmb_2d}:
the dashed line in that plot corresponds to
\begin{equation}
\Delta h \simeq -0.13 \, \left( \frac{\Delta M_\nu}{1~\mathrm{eV}} \right)~.
\label{cmb:eq:mnu-h2}
\end{equation}
The explanation for this mismatch is simple. Equation~(\ref{cmb:eq:mnu-h1}) assumed fixed $\theta_s$ {\it and} $\omega_{\rm cdm}$ values.
If instead we try to keep $\theta_s$ fixed while varying $\omega_{\rm cdm}$ according to equation~(\ref{cmb:eq:mnu-cdm}), we see increased correlation between $M_\nu$ and $h$, as shown by equation~(\ref{cmb:eq:mnu-h2})\footnote{Note that we estimated the correlation factor in equation~(\ref{cmb:eq:mnu-cdm}) with one significant digit, and in equation~(\ref{cmb:eq:mnu-h2}) with two significant digits: this is consistent with the fact that the correlation is much more clear and pronounced in the second case (the ratio of the minor over major axis is much smaller).}.

Hence, with CMB data only, the clearest and most important degeneracies involving the summed neutrino mass are between $M_\nu$ and $\omega_{\rm cdm}$ (due to lensing) and $M_\nu$ and $h$ (due to the angular diameter distance). There are other correlations, but they are much less pronounced. The third one would be between $M_\nu$ and $n_s$~\cite{Gerbino:2016sgw}. This can be understood by looking closely at the dotted-dashed in figure~\ref{fig:difftt} (lower right plot). The variation of $\omega_\mathrm{cdm}$ did not only rescale the amplitude of the CMB lensing potential, it also generated a small positive tilt.
The reason is that we have decreased the ratio $\omega_{\rm b}/\omega_{\rm cdm}$, thus changing the shape parameter controlling the effective spectral index of the matter power spectrum $P(k)$ for $k > k_{\rm eq}$: a smaller baryon amount relative to CDM implies a bluer spectrum.
Hence, the ($M_\nu$, $\omega_{\rm cdm}$) degeneracy is more pronounced when it goes together with a tiny decrease of the tilt $n_s$, by such a small amount that it would not conflict with temperature and polarisation data. This negative correlation is visible in Figure~\ref{fig:cmb_2d}, upper right plot.

\section{Effect of neutrino mass on the BAO scale}
\label{sec:bao}
The acoustic oscillations of the baryon-photon fluid that we observe in the CMB power spectrum produce a characteristic feature in the two point correlation function. In Fourier space the feature is located at a peculiar scale, the BAO scale, $k_\mathrm{BAO}=2\pi/r_s(z_{\rm drag})$, where $r_s(z_{\rm drag})$ is the comoving sound horizon at baryon drag
\begin{equation*}
r_s(z_{\rm drag})= \int_{0}^{\tau_{\rm drag}} c_s d\tau = \int_{z_{\rm drag}}^{\infty} \frac{c_s}{H(z)}dz.
\end{equation*}
The observed scale, assuming an isotropic fit of a galaxy sample\footnote{Anisotropic fit allow to disentangle the longitudinal information (i.e. the radial scale $H r_s$) from the transverse one (i.e. the tangential scale $D_A/r_s$).}, provides the ratio $r_s(z_{\rm drag})/D_V(z_{\rm BAO})$,
where $D_V$ is the volume distance, defined as
\begin{equation*}
D_V(z)=  \left[ z / H(z) (1+z)^2 d_A(z)^2 \right]^{1/3},
\end{equation*}
and $D_A=(1+z)d_A(z)$ is the comoving angular diameter distance. In the $\Lambda$CDM model with massive neutrinos, the ratio $r_s(z_{\rm drag})/D_V(z_{\rm BAO})$ can only depend on the four parameters $\{\omega_{\rm b}, \omega_{\rm cdm}, \omega_\nu, h\}$. More precisely, $r_s(z_{\rm drag})$ depends on the three parameters $\{\omega_{\rm b}, \omega_{\rm cdm}, h^2\}$,
while for redshifts below the non-relativistic transition, $z \ll z_{\rm nr} \sim 2 \times 10^3 (m_\nu/1 \,{\rm eV})$, $D_A(z)$ depends only on $\omega_{\rm tot} = \omega_{\rm b}+\omega_{\rm cdm}+\omega_\nu$ and  on $h$, because it can be approximated as
\begin{equation}
D_A(z) = \int_{0}^{z} \frac{ c dz'}{H(z')} \simeq 3000 \int_{0}^{z} \frac{dz'}{\sqrt{\omega_{\rm tot}(1+z')^3 + (h^2-\omega_{\rm tot})}}\,{\rm Mpc}.
\label{eq:DA}
\end{equation}
Note that the term inside the square root is a polynomial in $z'$ in which the constant term is precisely $h^2$ (so as expected, for small redshifts $z\ll1$, $D_A(z)$ depends {\it only} on the $h$ parameter, like in a Hubble diagram).

\begin{figure*}[h]
\centering
\includegraphics*[width=0.7\linewidth]{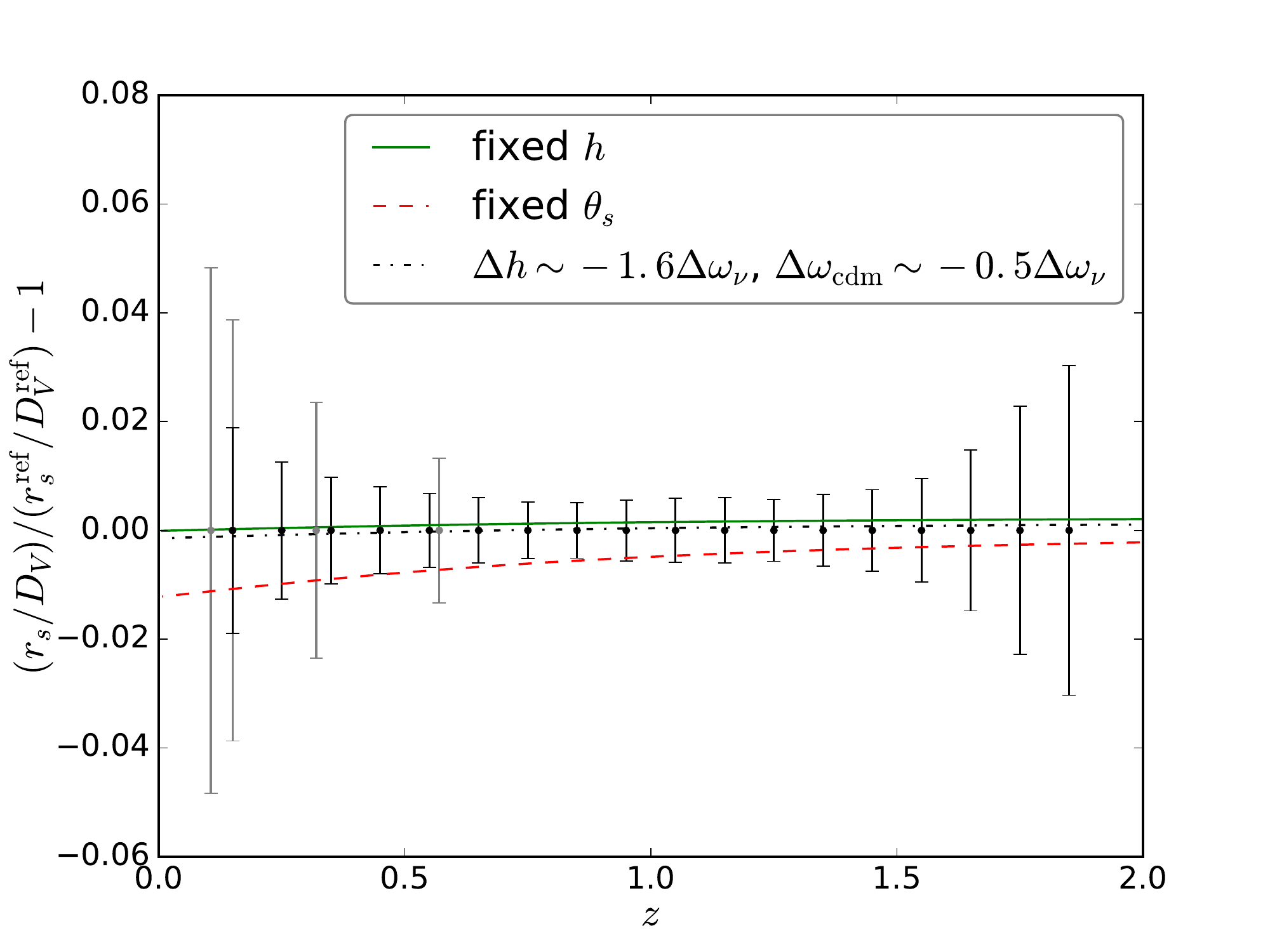}
\caption{ \label{fig:diffrsDV}
Relative error on $r_s/D_V$.
Gray error bars refer to the current BAO measurements: from left to right 6dFGRS~\cite{Beutler:2011hx}, SDSS MGS~\cite{Ross:2014qpa}, LOW-Z, C-MASS~\cite{Anderson:2013zyy}.
Black error bars mark the expected sensitivity of the future DESI experiment~\cite{Allison:2015qca,Font-Ribera:2013rwa}. Green solid line and red dashed lines are the same as in figure~\ref{fig:difftt}, i.e. higher $M_\nu$ with fixed $h$ (green solid line) and higher $M_\nu$ with fixed $\theta_s$ and varying $h$ (red dashed line). However, here the black dot dashed line is obtained by increasing $M_\nu$ and varying $h$ and $\omega_{\rm cdm}$ as in equations~(\ref{eq:corr_BAO}).
}
\end{figure*}
In figure~\ref{fig:diffrsDV} we show the residuals of current and future BAO measurements, taking as a reference the same model as before with $M_\nu = 60$~meV, as well as the relative difference on $r_s(z_{\rm drag})/D_V(z_{\rm BAO})$ between several models with a higher mass $M_\nu = 150$~meV (already introduced in section~\ref{sec:cmb}) and the reference model. For future measurement we take the example of DESI, assuming the same sensitivity as in Refs.~\cite{Allison:2015qca,Font-Ribera:2013rwa}.

We first vary only $M_\nu$ with fixed $\left\lbrace \omega_b, \omega_{\rm cdm}, h,n_s, A_s,\tau_{\rm reio}\right\rbrace$ (green solid line). This means that the early cosmological evolution is identical, while the matter density is slightly enhanced at late times (after the neutrino non-relativistic transition), by about one percent. Thus $d_s(z_{\rm dec})$ and $r_s(z_{\rm drag})$ are fixed, but 
$d_A(z_{\rm dec})$, $d_A(z_{\rm BAO})$ and $D_V(z_{\rm BAO})$ are subject to change. We have seen that this transformation shifted the CMB peaks by a detectable amount. However, the accuracy with which CORE will measure $\theta_s$ ($<0.01$\%) is much greater than that with which DESI will measure the BAO angular scales ($\sim$~1\%). From equation~\ref{eq:DA} we can see analytically that the typical variation of $D_A(z)$ between the two models is negligible for $z\ll 1$ and of the order of $\frac{1}{2} \frac{\Delta \omega_\nu}{\omega_{\rm tot}}\simeq 0.25\%$ for $1 < z < z_{\rm nr}$.  This explains why the green curve in figure~\ref{fig:diffrsDV} remains within the BAO error bars.

\begin{figure*}[h]
\begin{tabular}{cc}
\includegraphics*[width=0.45\linewidth]{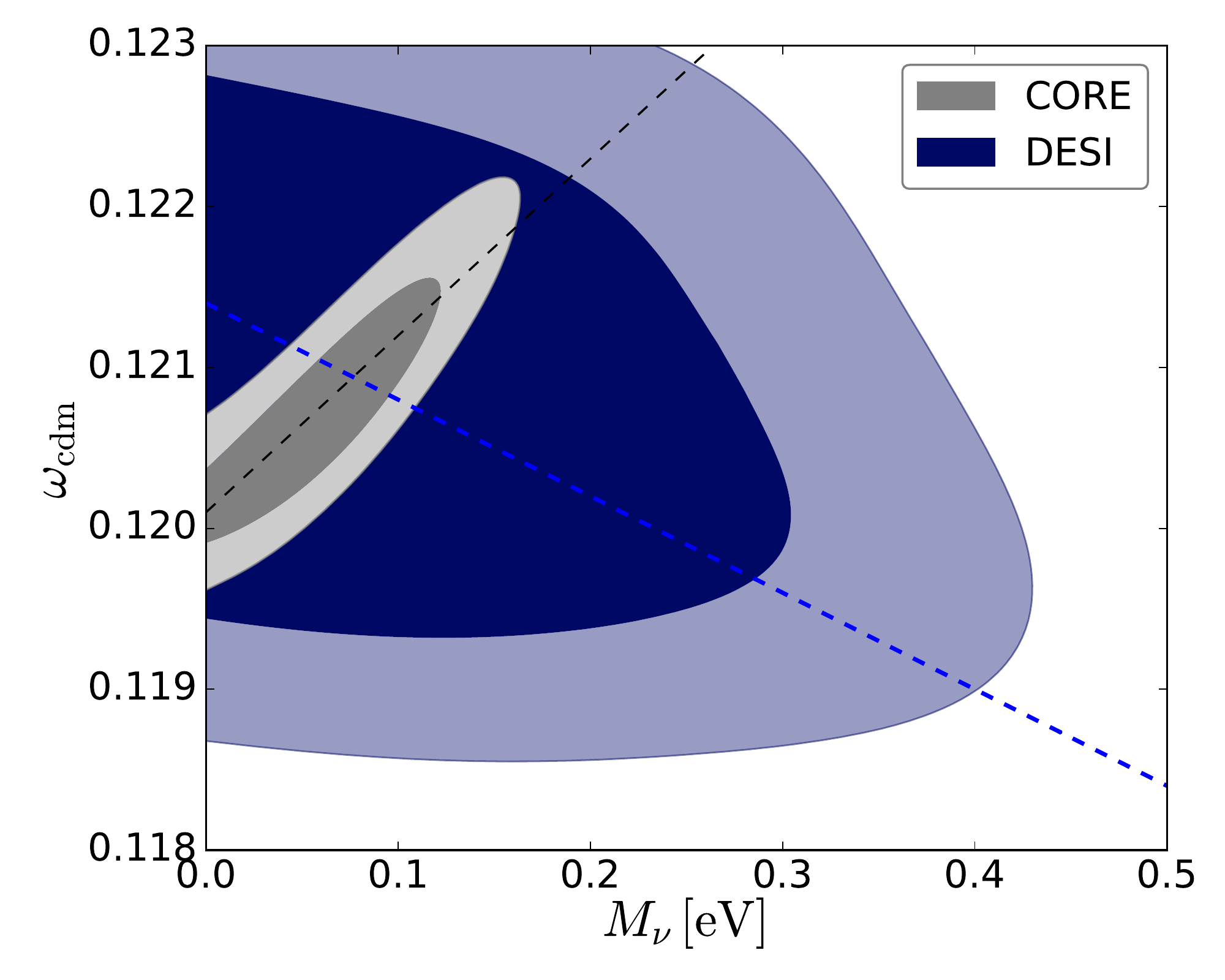}&
\includegraphics*[width=0.45\linewidth]{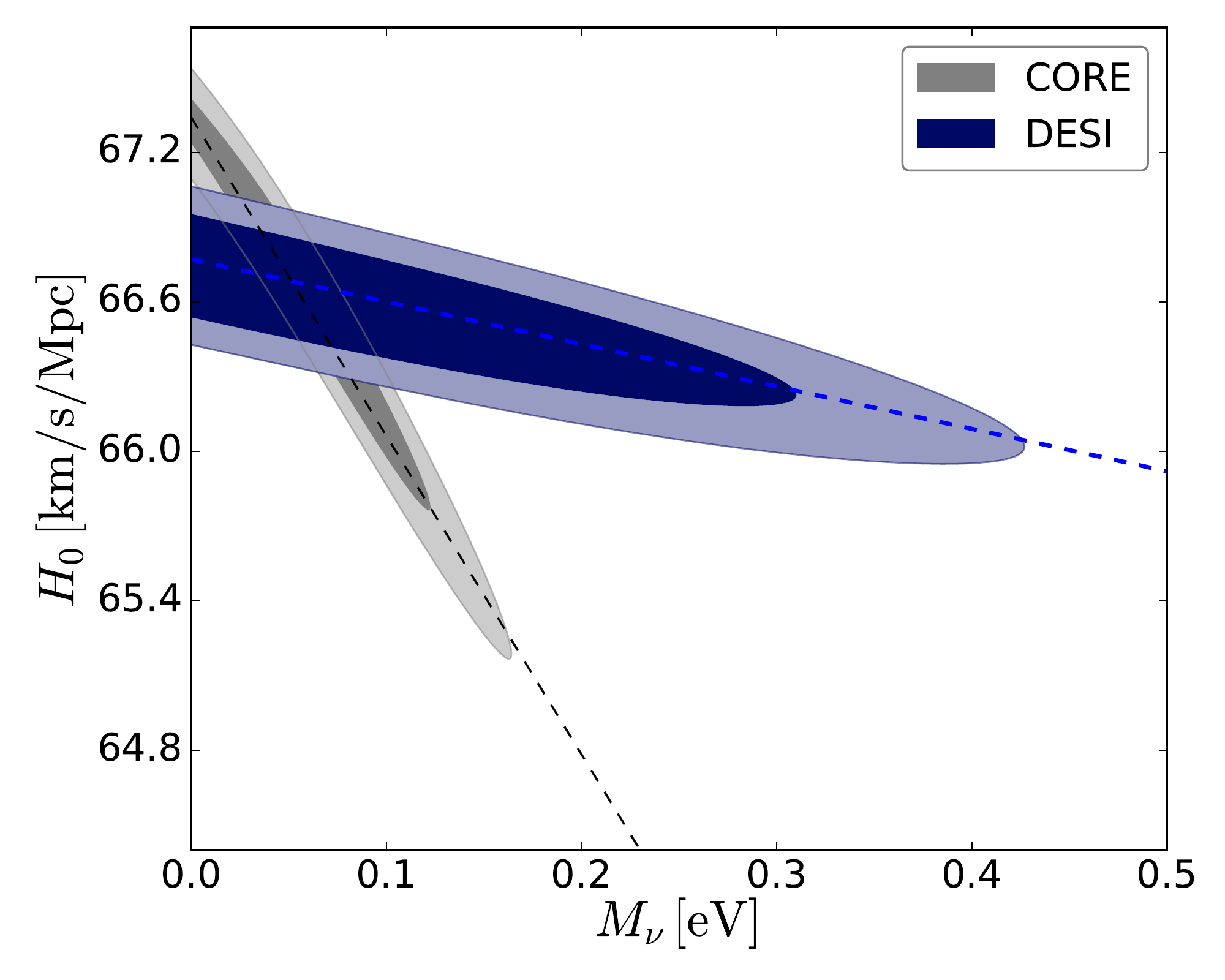} \\
\end{tabular}
\caption{ \label{fig:desionly_2d}
Marginalized one- and two- $\sigma$ contours in the plane $\left( \omega_{\rm cdm}, M_\nu \right)$ (left panel) and $\left(H_0, M_\nu \right)$ (right panel), for CMB-CORE or BAO-DESI mock data. The black dashed lines show the directions of degeneracy given in equations~(\ref{eq:corr_CMB}), and the blue ones in equations~(\ref{eq:corr_BAO}).}
\end{figure*}

\begin{figure*}[h]
\begin{tabular}{cc}
\includegraphics*[width=0.5\linewidth]{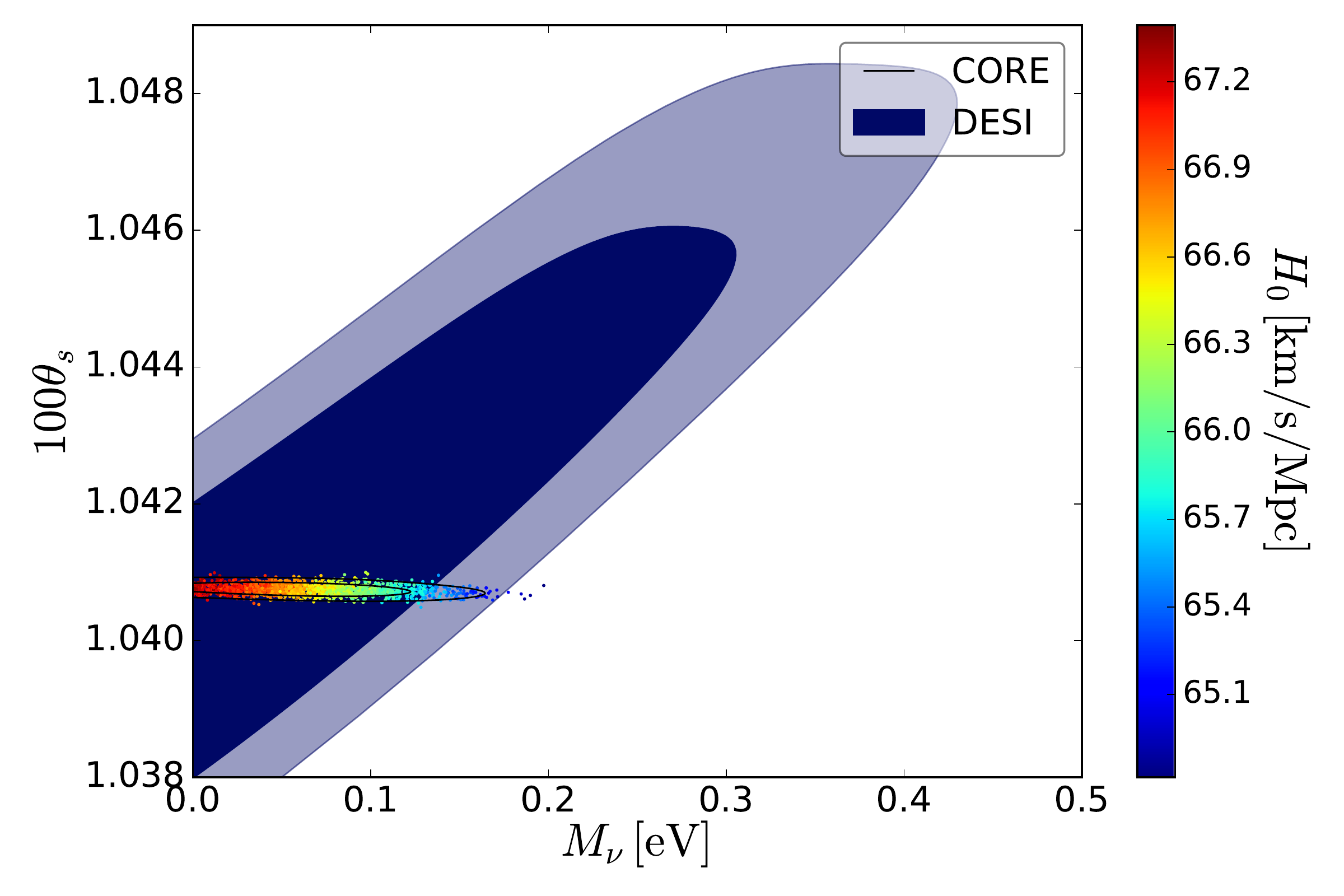} &
\includegraphics*[width=0.5\linewidth]{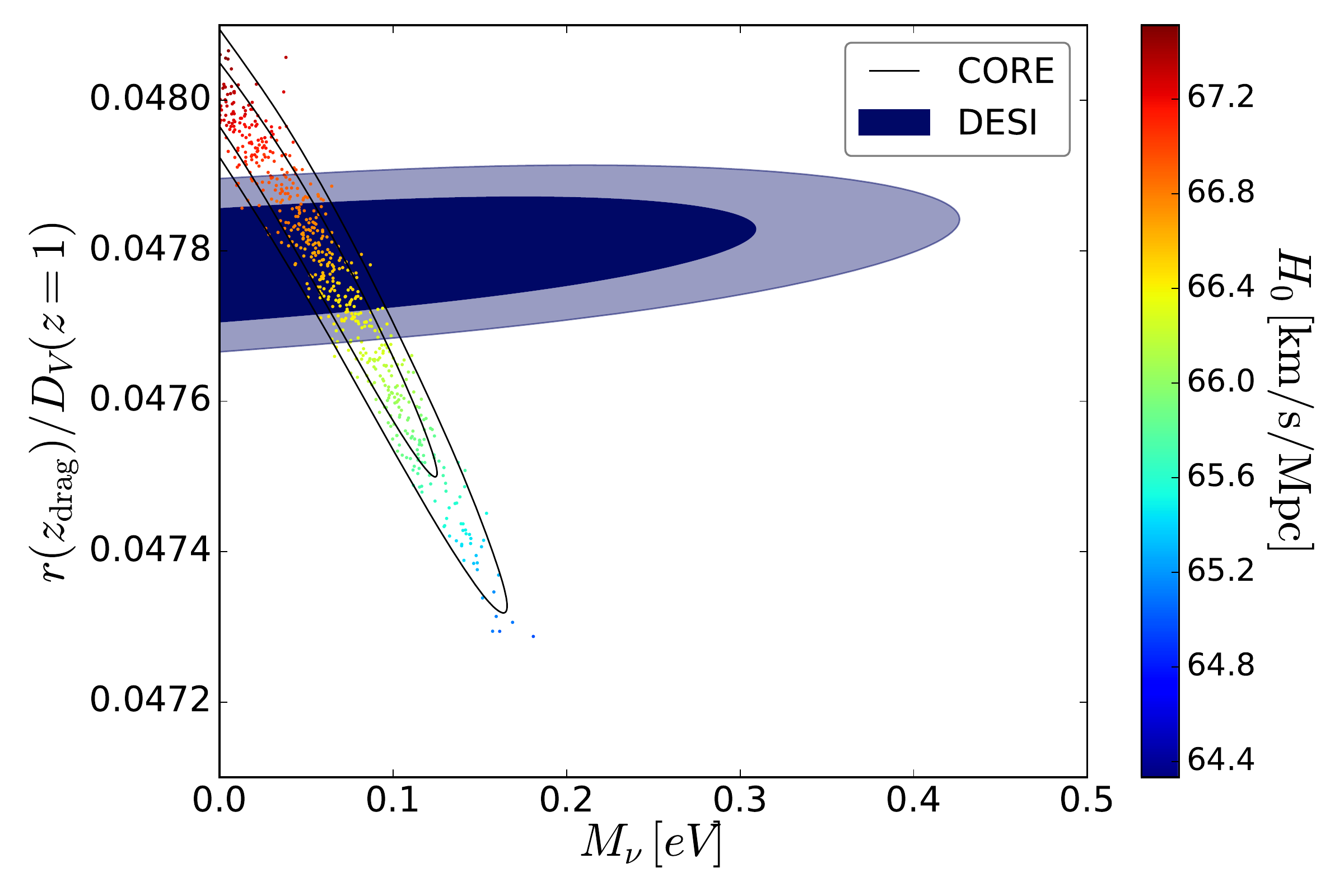} \\
\end{tabular}
\caption{ \label{fig:desionly_3d}
Marginalized one- and two- $\sigma$ contours in the plane $\left( \theta_s(z_{\rm dec}), M_\nu \right)$ (left) and $\left(r(z_\mathrm{drag})/D_V(z=1), M_\nu \right)$ (right),
for CMB-CORE or BAO-DESI mock data. In the CORE contours, samples are coloured according to the value of $H_0$.}
\end{figure*}

This preliminary discussion brings us to the key points of this section:
\begin{itemize}
\item {\it the BAO data alone can bound the neutrino mass, but not with great accuracy.} We showed previously that increasing $M_\nu$ with fixed $\left\lbrace \omega_b, \omega_{\rm cdm}, h\right\rbrace$ had no detectable effects, but this was because the mass variation was too small. If one keeps increasing $\omega_\nu$ with the other parameters fixed, the function inside the square root in equation~\ref{eq:DA} keeps increasing for the same value $z'$, and  $D_A(z)$ decreases. To avoid a BAO bound on $M_\nu$, one could try to exactly compensate the variation $\Delta \omega_\nu$ by an opposite variation in either $\omega_{\rm b}$ or $\omega_{\rm cdm}$, to keep $D_A(z)$ exactly constant. But in that case, the early cosmological evolution would change (sound speed, redshift of equality, redshift of baryon drag) and the ratio $r_s(z_{\rm drag})/D_V(z_{\rm BAO})$ would be shifted anyway. Hence there is no parameter degeneracy cancelling exactly the effect of $M_\nu$ in BAO observables, at least in the $\Lambda$CDM+$M_\nu$ model. This explains why in figures \ref{fig:desionly_2d} and \ref{fig:desionly_3d}, the contours involving $M_\nu$ are closed for DESI data alone, setting an upper bound on the summed mass of a few hundreds of meV.
\item {\it the strong degeneracy between $M_\nu$ and $h$ observed in the CMB case cannot exist with BAO data.} This denegeracy came from the possibility to keep constant angular scales ($\theta_s(z_{\rm dec})$, $\theta_d(z_{\rm dec})$) by varying $h$ with fixed $\left\lbrace \omega_b, \omega_{\rm cdm}\right\rbrace$. Indeed, when fitting CMB data with different neutrino masses, one can keep the same value of $d_A(z_{\rm dec})$ by altering the late time cosmological evolution: while $M_\nu$ tends to enhance the density at late times, one can decrease $h$ and the cosmological constant in order to compensate for this effect. This cannot be done with BAO data, because they probe $d_A(z)$ at several small values of $z$, comparable to the redshift of the transition $z_\Lambda$. The proof is particularly obvious if we look at equation~\ref{eq:DA} again. Whatever change in $h$ modifies the constant term inside the square root, and thus the value of $D_A(z)$ for $z \leq 1$. Thus the $(M_\nu,h)$ degeneracy discussed in the CMB section must be broken by BAO data. We get a first confirmation of this by looking at the red dashed curve in figure \ref{fig:diffrsDV}, obtained by increasing $M_\nu$ with a constant $\theta_s(z_{\rm dec})$: the new model departs from the other one by a detectable amount, at least given BAO-DESI errors (especially at $z\ll1$, as expected from this discussion). The second confirmation comes from the right plot in figure \ref{fig:desionly_2d}, showing very different correlations between 
$M_\nu$ and $h$ for CMB-CORE alone and BAO-DESI alone.
\item {\it there exists, however, a correlation between $M_\nu$, $h$ and $\omega_{\rm cdm}$ with BAO data, but along different angles than with CMB data.} This comes from the possibility to modify parameters in such a way that both $r_s(z_{\rm drag})$ and $D_V(z_{\rm BAO})$ get shifted, but almost by the same relative amount. To compensate for the effect of an increasing $\omega_\nu$, one has three parameters to play with: $\left\lbrace \omega_b, \omega_{\rm cdm}, h \right\rbrace$. However, $\omega_b$ is precisely fixed by CMB data alone, and for that reason we keep it to its Planck best-fit value. We then find that variations of the other two parameters by approximately
\begin{equation}
\Delta \omega_{\rm cdm} \sim - 0.5 \Delta \omega_\nu~, \qquad \Delta h \simeq -0.017 \left(\frac{\Delta M_\nu}{1~{\rm eV}}\right)  \simeq -1.6 \, \Delta \omega_\nu 
\label{eq:corr_BAO}
\end{equation}
achieve a nearly constant ratio $r_s(z_{\rm drag})/D_V(z_{\rm BAO})$ in the redshift range best probed by the BAO-DESI experiment. As argued before, this ratio is more sensitive to $h$ than $\omega_{\rm cdm}$ in that range, so the correlation between $\omega_{\rm cdm}$ and $\omega_\nu$ is weak, while that between $h$ and $\omega_\nu$ is strong (see Figure~\ref{fig:desionly_2d}).
\end{itemize}
The parameter correlations found in eq.~(\ref{eq:corr_BAO}) for BAO data are very different from those found in the previous section for CMB data:
\begin{equation}
\Delta \omega_{\rm cdm} \sim  \Delta \omega_\nu~, \qquad \Delta h \simeq -0.13 \left(\frac{\Delta M_\nu}{1~{\rm eV}}\right)  \simeq -12 \, \Delta \omega_\nu~.
\label{eq:corr_CMB}
\end{equation}
The combination of CMB and BAO data can thus break these degeneracies, as it is often the case when combining high and low redshift probes of the expansion history. The breaking does not arise from the joint measurement of $\omega_{\rm cdm}$ and $\omega_\nu$ (because BAO data are much less sensitive to $\omega_{\rm cdm}$ alone than CMB data), but from that of $h$ and $\omega_\nu$, for which the different directions of degeneracy appear very clearly on figure~\ref{fig:desionly_2d}.
Thus, the future BAO-DESI data will contribute to tighter constraints on $M_\nu$.


Another way to illustrate the degeneracies discussed here is to fit CMB data or BAO data alone with a $\Lambda$CDM+$M_\nu$ model, and to plot the results in the space of parameters $(M_\nu, \theta_s(z_{\rm dec}))$ and $(M_\nu,   r_s(z_{\rm drag})/D_V(z_{\rm BAO}))$ for a median redshift $z_{\rm BAO}=1$. This is shown in figure~\ref{fig:desionly_3d}. When fitting CMB alone, thanks to the degeneracy of equations~(\ref{eq:corr_CMB}), we can increase $M_\nu$ while keeping $d_A(z_{\rm dec})$ and $\theta_s(z_{\rm dec})$ fixed (left plot), but this is at the expense of decreasing the BAO angular scale by more than allowed by observational errors (right plot). Conversely, when fitting BAO data alone, we can play with the degeneracy of equations~(\ref{eq:corr_BAO}) to keep the BAO angular scale fixed, but this requires $\theta_s(z_{\rm dec})$ to vary. The right plot in figure~\ref{fig:desionly_3d} illustrates, in an alternative way to the right plot of figure~\ref{fig:desionly_2d}, how the combination of the two data sets can improve neutrino mass bounds.

Finally, we expect, as a secondary indirect effect, that the correlation between $M_\nu$ and $\left(A_s, \tau_{\rm reio}\right)$ will be more noticeable in a combined analysis of CMB and BAO than for CMB alone. In section~\ref{sec:cmb}, we mentioned in points 3 and 4 that the impact of $M_\nu$ on CMB lensing could be compensated in two ways: by increasing either $\left(A_s, \tau_{\rm reio}\right)$ (point 3) or $\omega_{\rm cdm}$ (point 4). We explained why the former option is favoured with CMB data alone. Since we just argued that BAO data can reduce the degeneracy between neutrino masses and $\omega_{\rm cdm}$, the latter option is more relevant  when the data are combined with each other. Indeed, we will see a small correlation between $\left(M_\nu, \tau_{\rm reio}\right)$ in the combined results presented in section~\ref{sec:results}, one that was hardly noticeable with CMB alone. Of course, this degeneracy is not perfect, and extends only up to the point at which $\tau_{\rm reio}$ becomes too large to be compatible with CMB polarisation data.
\section{Effect of neutrino mass on Large Scale Structure observables}
\label{sec:lss}

\subsection{Cosmic shear and galaxy clustering spectrum}
\label{sec:lss-spec}

The Euclid satellite, whose launch is scheduled for 2020, will provide the most accurate ever galaxy redshift survey, measuring cosmological observables, such as cosmic shear and galaxy clustering, with 1\% accuracy. Euclid data will certainly lead to a major breakthrough in precision cosmology thanks to very precise low redshift measurement which will break the CMB degeneracies among cosmological parameters (see references~\cite{Carbone:2010ik,Namikawa:2010re,Hamann:2012fe,Audren:2012vy,DiDio:2013bqa,DiDio:2013sea,Basse:2013zua,Basse:2014qqa,Archidiacono:2015mda}).
Here we use the information extracted from the cosmic shear power spectrum projected in angular harmonics (2D) and the galaxy clustering power spectrum (3D). Both observable are related to the non-linear matter power spectrum depending on wavenumber and redshift, $P_\mathrm{m}(k,z)$. In our forecasts, we estimate this quantity using the {\sc halofit} algorithm, updated by \cite{Takahashi:2012em} and also by \cite{Bird:2011rb} for the effect of neutrino masses, as implemented in {\sc class} v2.5.0.

\vspace{0.2cm}

\noindent {\it Cosmic shear.}
The cosmic shear auto and cross correlation angular power spectrum in the $i$ and $j$ redshift bins is given in the Limber approximation by:
\begin{equation}
C_\ell^{ij}=H_0^4 \int^{\infty}_{0} \frac{dz}{H(z)} W_i(z) W_j(z) P_{\rm m} \left( k=\frac{l}{r(z)}, z\right),
\label{eq:Cl}
\end{equation}
where the window functions are given by
\begin{equation}
W_i(z)=\frac{3}{2} \Omega_m (1+z) \int^{\infty}_{0}dz_s \frac{n_i(z_s)(r(z_s)-r(z))}{r(z_s)},
\label{eq:W}
\end{equation}
and the number of galaxies per steradian in the $i$ bin is given by
\begin{equation*}
n_i(z)=\frac{\int_{z_i^{\rm min}}^{z_i^{\rm MAX}} dn/dz \mathcal{P}(z,z_\mathrm{ph})dz_\mathrm{ph}}
{\int_{0}^{\infty} dn/dz \mathcal{P}(z,z_\mathrm{ph})dz_\mathrm{ph}}
\end{equation*}
with $ \mathcal{P}(z,z_\mathrm{ph})$ being the error function
\begin{equation*}
\mathcal{P}(z,z_\mathrm{ph})=\frac{1}{\sqrt{2 \pi \sigma_\mathrm{ph}^2}} \exp \left[ - \frac{1}{2} \left( \frac{z-z_\mathrm{ph}}{\sigma_\mathrm{ph}}\right)\right].
\end{equation*}
We use the Euclid prescription for the galaxy surface density
\begin{equation*}
dn/dz=z^2 \exp \left[ -(z/z_0)^{1.5} \right]
\end{equation*}
with $z_\mathrm{mean}=1.412z_0$.
Finally we consider a photometric redshift error $\sigma_\mathrm{ph}=0.05(1+z)$, sky fraction $f_\mathrm{sky}=0.3636$, mean internal ellipticity $0.22$ and total number of observed galaxies $30$ per arcmin$^2$.

\vspace{0.2cm}

\noindent {\it Galaxy clustering.} For galaxy clustering the observed power spectrum reads:
\begin{equation*}
P(k_\mathrm{ref},\mu,z) = \frac{D_A(z)^2_\mathrm{ref} H(z)}{D_A(z)^2H(z)_\mathrm{ref}} b(z)^2 \left[ 1 + \beta(z,k(k_\mathrm{ref},\mu,z)) \mu^2 \right]^2
\times P_\mathrm{m}(k(k_\mathrm{ref},\mu,z)) e^{-k(k_\mathrm{ref},\mu,z)^2,z) \mu^2 \sigma_r^2},
\label{eq:Pg}
\end{equation*}
where $\mu$ is the cosine of the angle between the line of sight and the wavenumber in the reference cosmology (ref) $k_\mathrm{ref}$, $k$ is the wavenumber in the true cosmology and it is defined as a function of $k_\mathrm{ref}$
\begin{equation*}
k^2=\left( \frac{(1-\mu^2)D_A(z)^2_\mathrm{ref}}{D_A(z)^2} + \frac{\mu^2H(z)^2}{H(z)_\mathrm{ref}^2} \right)k_\mathrm{ref}^2.
\end{equation*}
The factor $\left[ D_A(z)^2_\mathrm{ref} H(z) \right] / \left[ D_A(z)^2H(z)_\mathrm{ref} \right]$ encodes the geometrical distortions related to the Alcock-Paczynski effect.
The bias can be written as $b=\sqrt{(1+z)}$, $\beta$ encodes the redshift space distortions
\begin{equation*}
\beta(k,z)=\frac{1}{2b(z)} \frac{d \ln P_\mathrm{m}(k,z)}{d \ln a},
\end{equation*}
and finally the spectroscopic redshift error is $\sigma_r=dr(z)/dz \sigma_z$.

Both the $C_\ell^{ij}$ and the $P(k)$ provide information on a broad range of scales; therefore, given the same survey sensitivity, they are more efficient than BAO in constraining cosmological parameters; however, for the very same reason, they are more prone to systematic effects such as residual errors in the estimate of non-linear corrections, non-linear light-to-mass bias or redshift space distortions (see e.g. \cite{Rassat:2008ja,Hamann:2010pw}). For that reason, we include in our forecast a theoretical error on the observable power spectrum, increasing above a given redshift-dependent scale of non-linearity~(see \cite{Audren:2012vy}, or \cite{Baldauf:2016sjb} for a more refined treatment). 

The assumed theoretical error amplitude has a direct impact on the galaxy clustering sensitivity to cosmological parameters.
Here we stick to the approach of  \cite{Audren:2012vy}, and we refer to this work for details and equations. As emphasised in \cite{Baldauf:2016sjb}, this approach is extremely (and maybe overly) conservative, because the error is assumed to be uncorrelated between different $k$-bins. The error grows as a function of the ratio $k/k_{\rm nl}(z)$, where
$k_{\rm nl}(z)$ is the redshift-dependent scale of non linearity, with a shape and amplitude inspired from the typical residuals between different N-body codes\footnote{The error function is explicitly given by $\alpha(k,z) \equiv \frac{\Delta P_{\rm m}(k,z)}{P_{\rm m}(k,z)} = \frac{ \ln \left[ 1 + k/k_{\rm nl}(z) \right]}{1 + \ln \left[ 1 + k/k_{\rm nl}(z) \right]} \epsilon$, 
where $k_{\rm nl}(z)$ is identical to the quantity $k_{\sigma}(z)$ computed at each redshift by {\sc Halofit}, and the error amplitude parameter $\epsilon$ is the unique free parameter in this model. The asymptotic error in the deep non-linear regime is then given by $100 \epsilon\,\%$ .}. Choosing a value for the error amplitude parameter $\epsilon$ amounts to estimating the accuracy of grids of N-body simulations and of models for various non-linear and systematic effects in a few years from now. The baseline choice in  \cite{Audren:2012vy} was $\epsilon=0.05$. In this paper, given the progress in the field observed since 2012, we choose to reduce it to  $\epsilon=0.025$, meaning that the uncorrelated theoretical error saturates at the 2.5\% level in the deep non-linear regime.
This error is explicitly shown in figure~\ref{fig:diffgc} for $z=0.5$ and $z=2$, and its impact on the lensing harmonic spectrum appears in figure~\ref{fig:diffll} for the lowest and highest redshift bins of the Euclid lensing survey. 
In presence of a theoretical error, the issue of where to cut the integrals in the galaxy clustering likelihood becomes hardly relevant, provided that the cut-off is chosen in the region where the theoretical error dominates.
In  what follows, we will cut the observable $P(k_\mathrm{ref},\mu,z)$ at $k_\mathrm{max}=0.6\, h/$Mpc for all redshifts. 
For cosmic shear, the inclusion of the theoretical error is also important, although the observational error bar does not decreases indefinitely with $\ell$ due to the finite angular resolution of the shear maps. In our forecasts, we perform a cut at $\ell_\mathrm{max}=2000$.


\subsection{Degeneracies between $M_\nu$ and other parameters}

\begin{figure*}[h]
\begin{tabular}{cc}
\includegraphics*[width=0.5\linewidth]{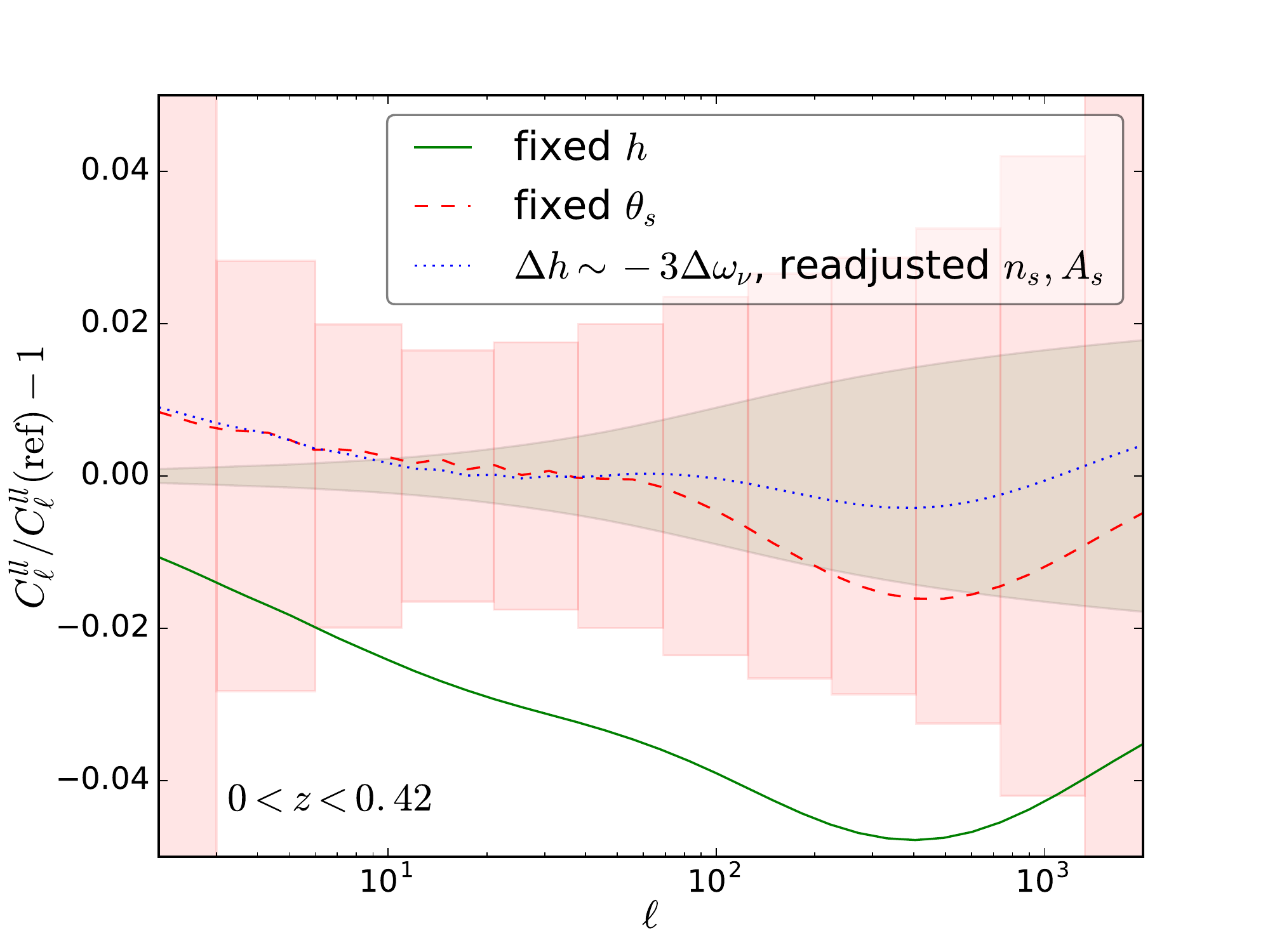} &
\includegraphics*[width=0.5\linewidth]{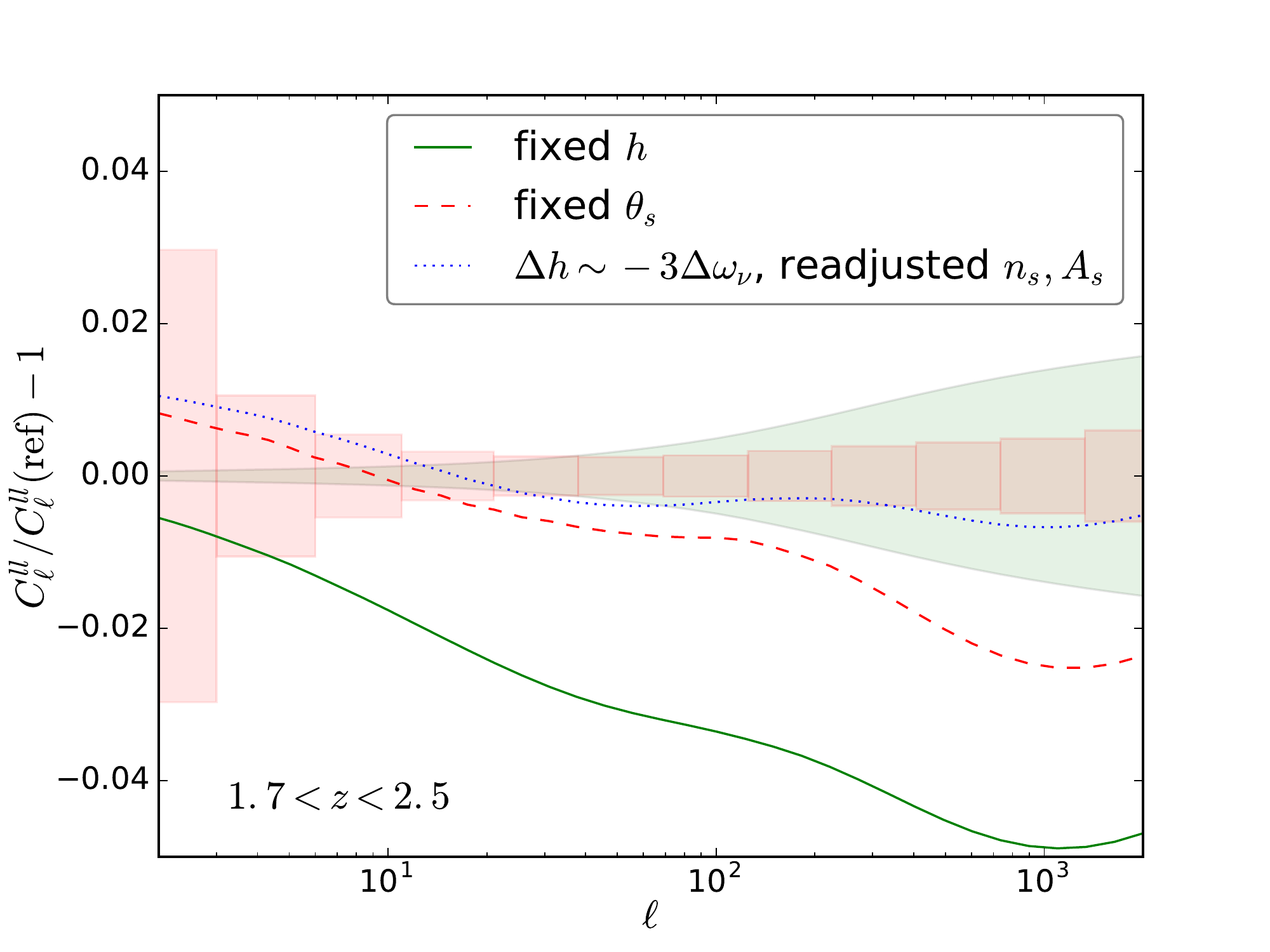} \\
\end{tabular}
\caption{ \label{fig:diffll}
Relative error on the galaxy lensing $C_\ell^{ll}$ in the first redshift bin ($0<z<0.42$, left panel) and in the tenth redshift bin ($1.7<z<2.5$, right panel). Here the redshift range is $0<z<2.5$ and is divided in ten equi-populated redshift bins. The light pink rectangles refers to the observational error. The light green shaded area shows the relative error associated to our model for the  theoretical uncertainty on $P_m(k,z)$.
Green solid and red dashed lines are the same as in figure~\ref{fig:difftt}, i.e. higher $M_\nu$ with fixed $h$ (green solid line) and higher $M_\nu$ with fixed $\theta_s$ and varying $h$ (red dashed line). The blue dotted line, besides the higher $M_\nu$, implies a smaller value of $h$ ($\Delta h \sim - 3 \Delta \omega_\nu$), an increase of $n_s$ by  0.4\% and of $A_s$ by 2\%.
}
\end{figure*}

\begin{figure*}[h]
\begin{tabular}{cc}
\includegraphics*[width=0.5\linewidth]{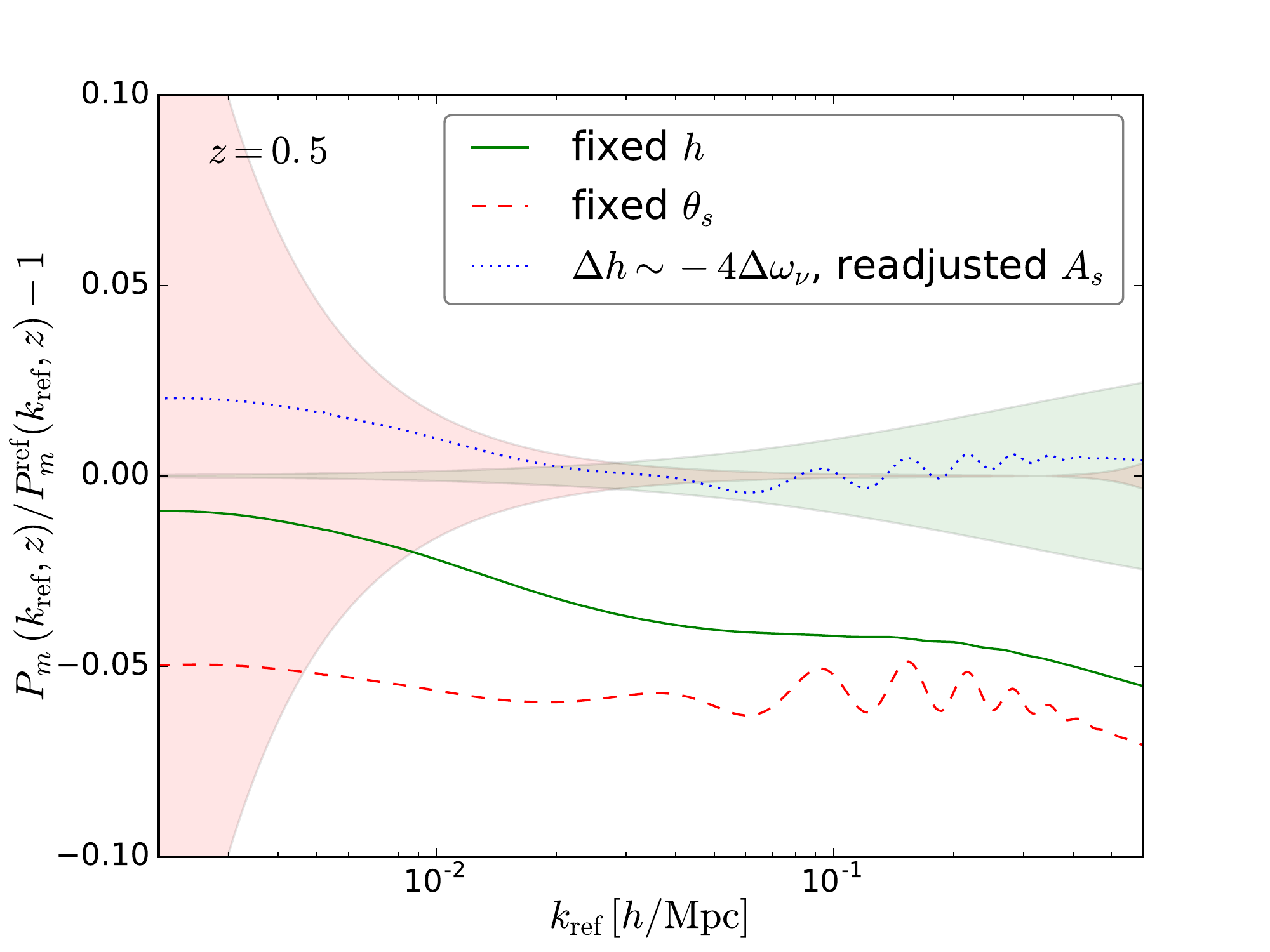} &
\includegraphics*[width=0.5\linewidth]{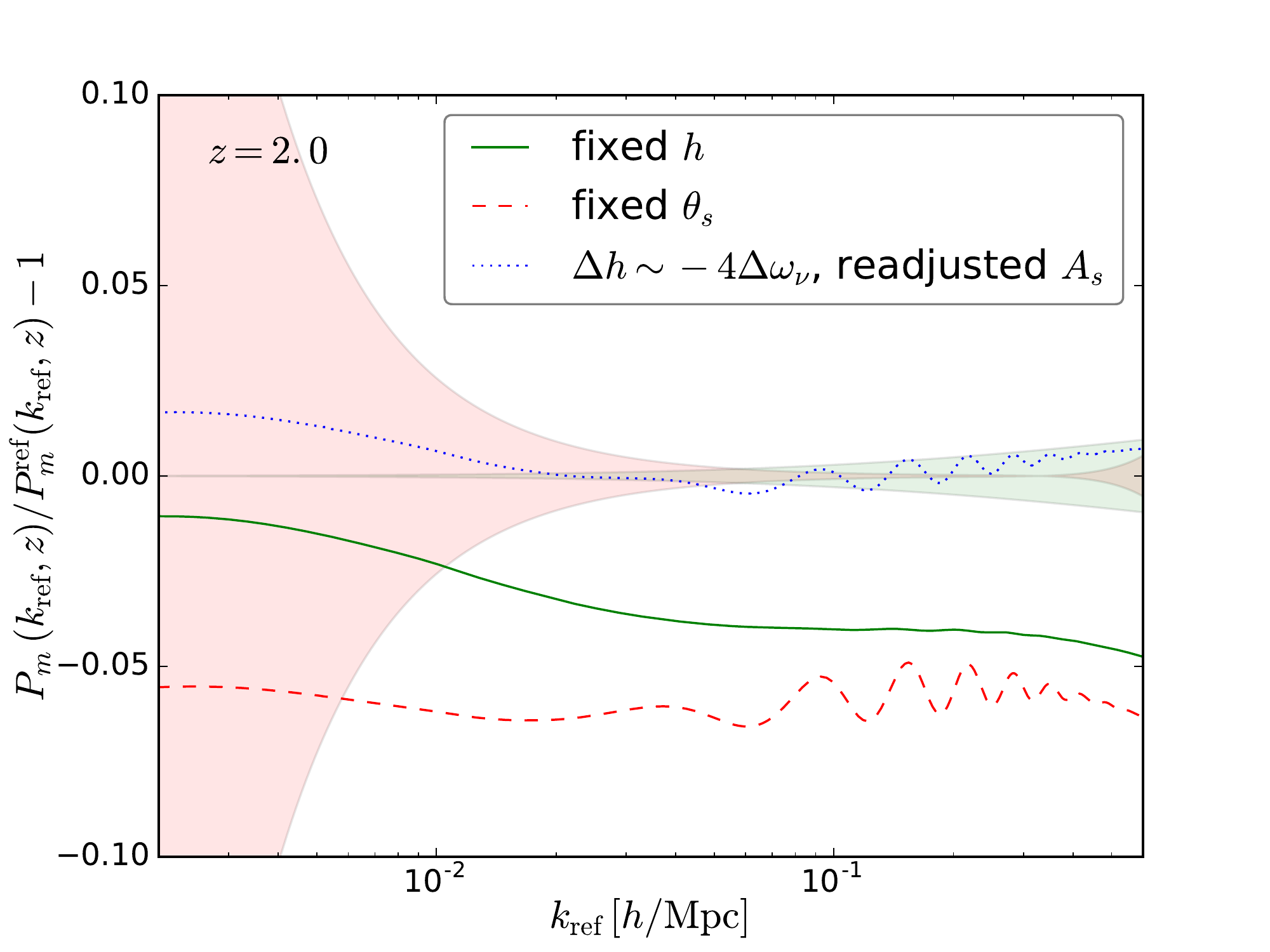} \\
\end{tabular}
\caption{ \label{fig:diffgc}
Relative error on the non linear matter power spectrum $P_{\rm m}(k_{\rm ref},z)$ perpendicular to the line of sight ($\mu = 0$) at redshift $z=0.5$ (left panel) and $z=2$ (right panel). The light pink shaded area refers to the observational error, including cosmic variance. The light green shaded area shows our model for the theoretical uncertainty.
Here the redshift range is $0.5<z<2$ and is divided in 16 redshift bins. 
Green solid and red dashed lines are the same as in figure~\ref{fig:difftt}, i.e. higher $M_\nu$ with fixed $h$ (green solid line) and higher $M_\nu$ with fixed $\theta_s$ and varying $h$ (red dashed line). The blue dotted line, besides the higher $M_\nu$, implies a smaller value of $h$ ($\Delta h \sim -4 \Delta \omega_\nu$) and an increase of $A_s$ by 5\%. 
}
\end{figure*}

In figure~\ref{fig:diffll} and \ref{fig:diffgc} we show the relative shift in the shear power spectrum and in the galaxy power spectrum that is obtained when increasing the summed neutrino mass while keeping various quantities fixed. We also show for comparison the observational and theoretical errors computed in the same way as in Ref.~\cite{Audren:2012vy}, using the survey specifications listed above. We will study the impact of increasing the summed neutrino mass on these observables: (1) when keeping the usual cosmological parameters fixed, (2) when tuning $h$ at the same time in order to keep the same angular peak scale in the CMB, and (3) when playing with other parameters in order to minimize the impact of neutrino mass on LSS observables. The discussion in (2) (respectively, (3)) is relevant for understanding the degeneracy between $M_\nu$ and other parameters when fitting CMB+LSS data (respectively, LSS data alone). 

\begin{figure*}[h]
\begin{tabular}{cc}
\includegraphics*[width=0.45\linewidth]{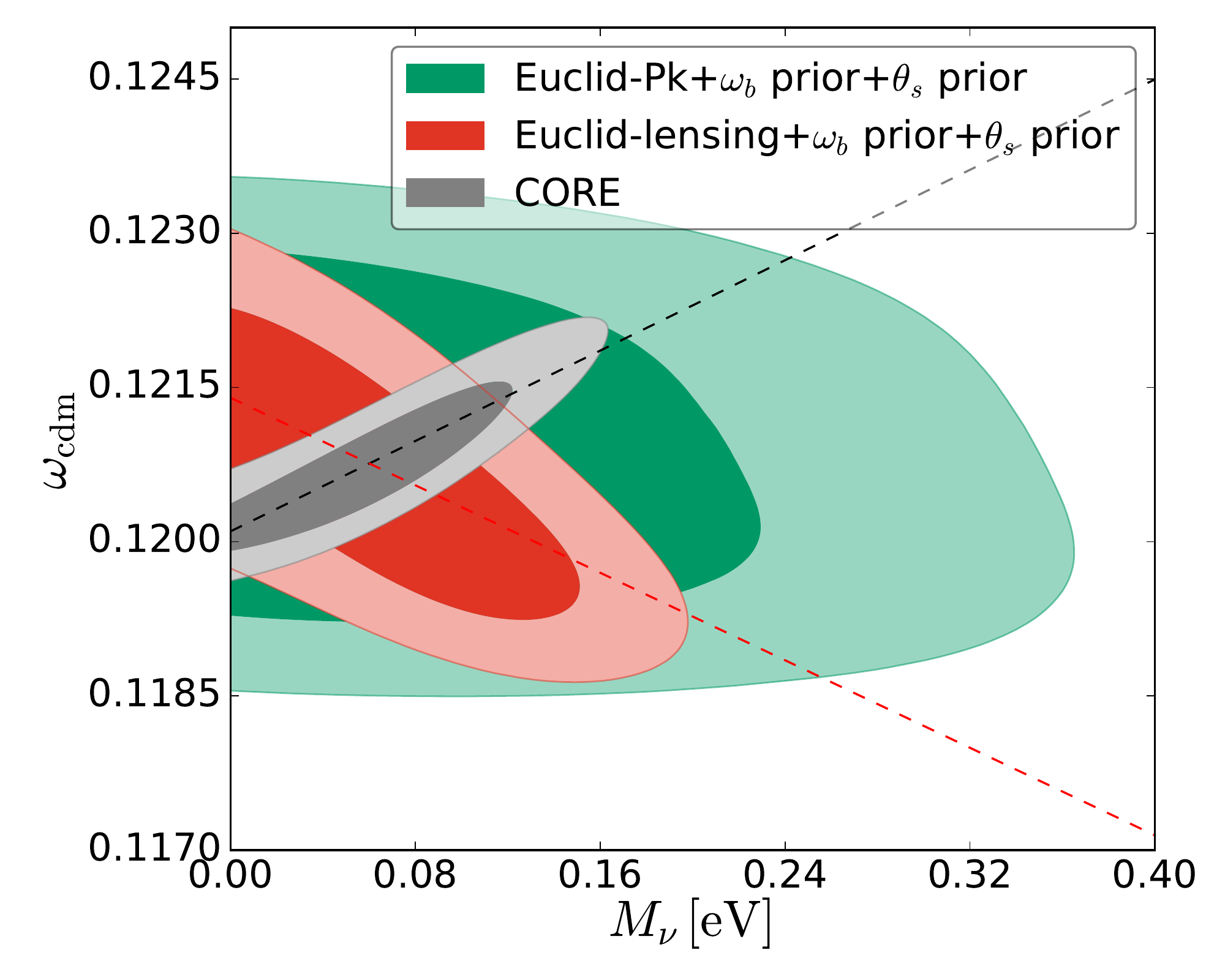} &
\includegraphics*[width=0.45\linewidth]{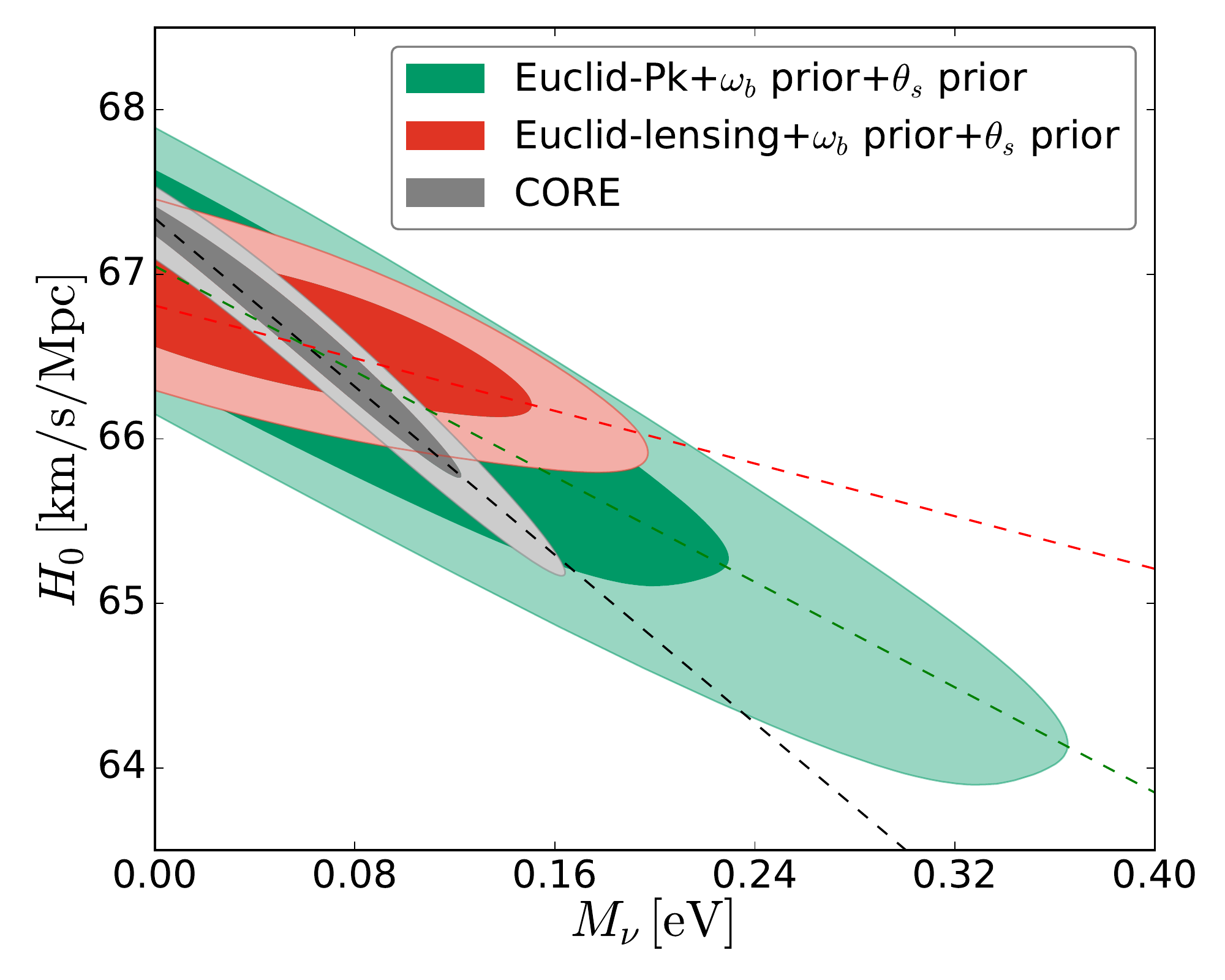} \\

\includegraphics*[width=0.45\linewidth]{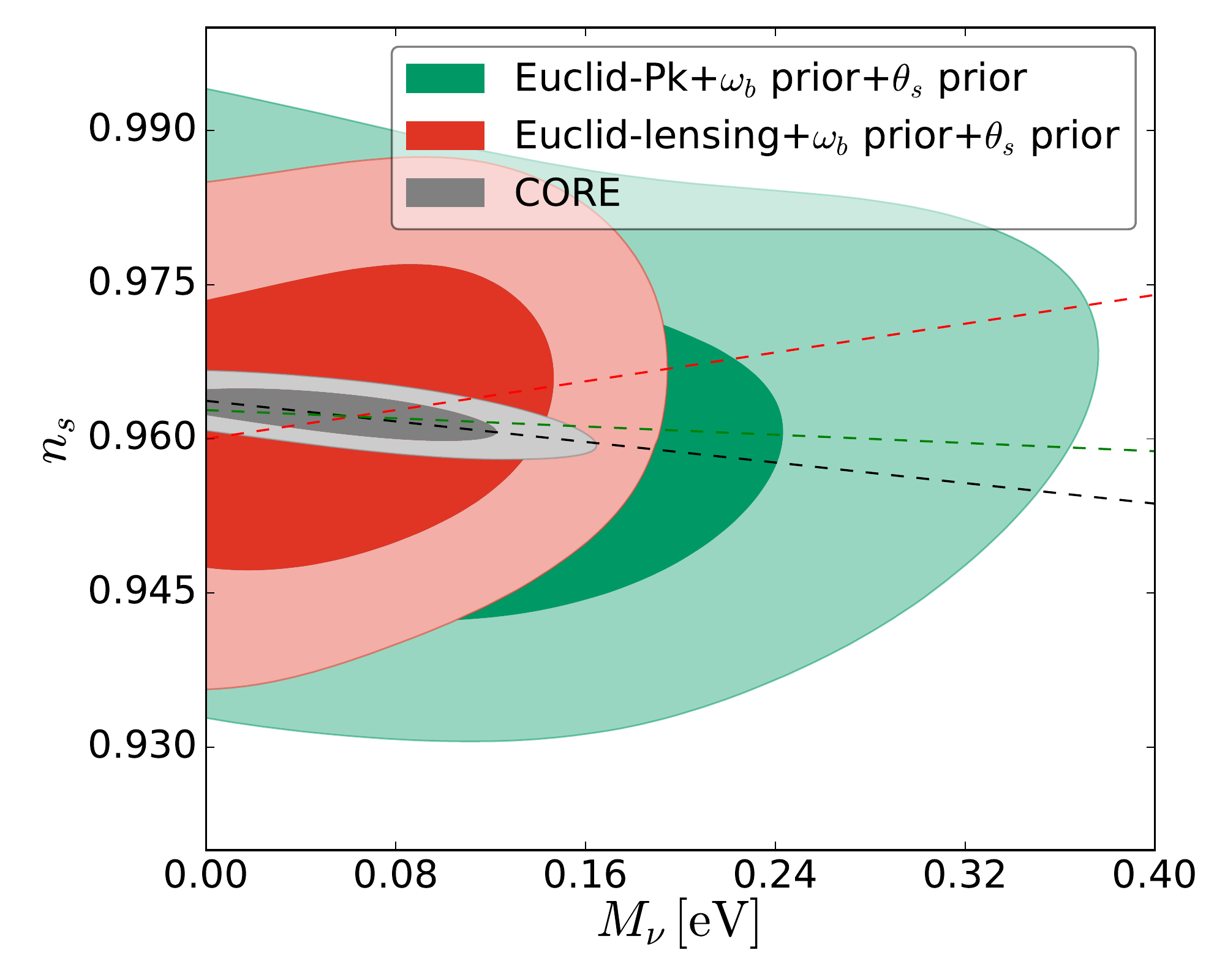} &
\includegraphics*[width=0.45\linewidth]{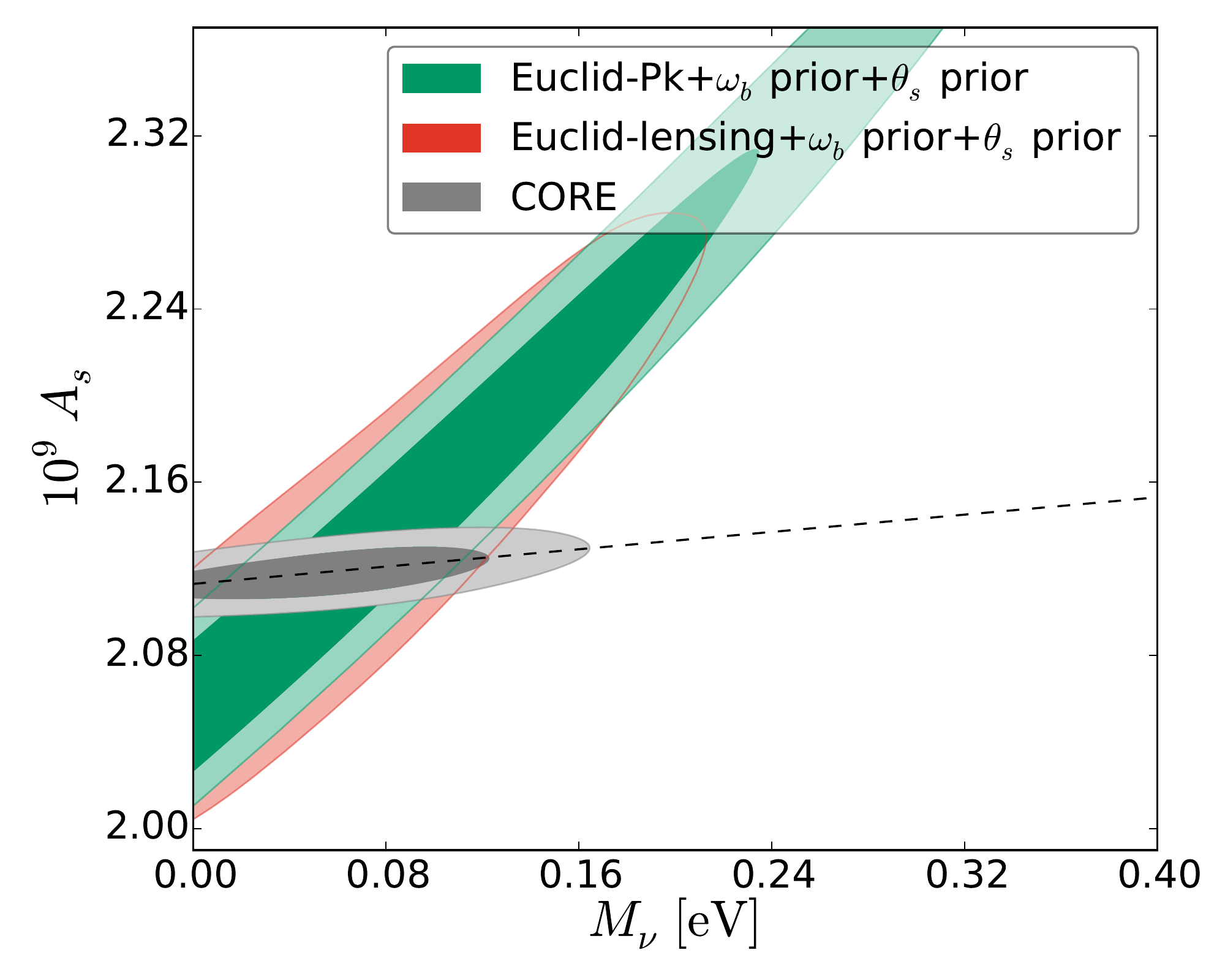}
\end{tabular}
\caption{ \label{fig:euclidonly}
Marginalized one- and two- $\sigma$ contours in the plane $\left(\omega_\mathrm{cdm}, M_\nu \right)$ (upper left panel), $\left(H_0, M_\nu \right)$ (upper right panel), $\left(n_s, M_\nu \right)$ (bottom left panel), $\left(A_s, M_\nu \right)$ (bottom right panel). The black dashed lines show the degeneracies encoded in CMB data, the red and green dashed lines account for some of the most prominent correlations arising from cosmic shear and galaxy clustering, respectively.
}
\end{figure*}

As in the previous sections, we will then check our theoretical conclusions through an MCMC forecast of the sensitivity of future experiments that will measure the spectra discussed above. In  figure~\ref{fig:euclidonly} we plot the marginalized one- and two- $\sigma$ contours showing the degeneracies at study:
 $\left(\omega_\mathrm{cdm}, M_\nu \right)$ (upper left panel), $\left(H_0, M_\nu \right)$ (upper right panel), $\left(n_s, M_\nu \right)$ (bottom left panel), $\left(A_s, M_\nu \right)$ (bottom right panel).
The CORE only contours (in gray) are the same as in figure~\ref{fig:desionly_2d}. The Euclid related contours have been obtained through an MCMC forecast including either galaxy clustering (in green) or cosmic shear (in red), following the specifications listed in section~\ref{sec:lss-spec}. Fitting Euclid mock data alone would return wide contours in parameter space. Given that the two quantities best measured by CMB experiments are the angular scale of the acoustic horizon and the baryon density, the question in which we are most interested is: assuming that information on $\omega_b$ and $\theta_s$ is provided by a CORE-like CMB experiment, what is the pull on other parameters coming from Euclid alone? To address this, when fitting Euclid data, we impose two uncorrelated gaussian priors on respectively $\omega_b$ and $\theta_s$, with standard deviations taken from our previous CORE-CMB forecast, while keeping $\tau_{\rm reio}$ fixed, since the latter does not affect galaxy clustering and shear observables in any way.


\begin{enumerate}
\item {\it Neutrino mass effects with all standard cosmological parameters fixed: the usual neutrino--induced step--like suppression.}

Like in the previous sections, we start by increasing the summed neutrino mass from $M_\nu=0.06$~eV to $M_\nu=0.15$~eV, keeping all the other cosmological parameters $\left\lbrace \omega_b, \omega_{\rm cdm}, h,n_s, A_s,\right\rbrace$ fixed (green solid line). Note that in most of the literature, the effect of neutrino masses on the matter power spectrum is discussed precisely in that way. One reason is that fixing $\left\lbrace \omega_b, \omega_{\rm cdm}, n_s, A_s,\right\rbrace$ amounts in keeping the same ``early cosmological evolution'' until the time of the neutrino non-relativistic transition. The choice to fix also $h$ is mainly a matter of simplicity.

As expected, the larger $M_\nu$ induces a relative suppression of power on small scales compared to large scales, visible both in the shear and in the galaxy power spectrum. To be precise, in the redshift range surveyed by Euclid, $0<z<2.5$, neutrinos with a mass $M_\nu > 0.05$~eV are already well inside the non-relativistic regime, thus, the spectrum is suppressed on scales smaller than the free-streaming scale $k>k_\mathrm{fs}(z)$. In the redshift range of interest, $0<z<2.5$, the free streaming wavenumber spans the range $[0.0077-0.0041]h\mathrm{Mpc}^{-1}$ (respectively, $[0.0192-0.0103]h\mathrm{Mpc}^{-1}$) for $M_\nu=0.06$~eV (respectively, $M_\nu=0.15$~eV)\footnote{The free streaming length depends on the mass of each neutrino rather than on the sum. Here we have assumed three massive degenerate neutrinos.}. The suppression in power makes both the $C_\ell^{ij}$ and the $P(k)$ directly sensitive to the neutrino mass sum, while this was not the case for the purely geometrical information encoded in BAO measurements. 

This sensitivity is reinforced by non-linear effects which are well visible on figures~\ref{fig:diffll} and \ref{fig:diffgc}. In the shear spectrum of figure~\ref{fig:diffll}, in absence of non-linear corrections, the green curve would be almost constant for $\ell>100$. Non-linear gravitational clustering produces a characteristic ``spoon shape'' or dip~\cite{Bird:2011rb}. The minimum of the dip is seen at $\ell \sim 40$ in the first redshift bin and $\ell \sim 1000$ in the last one. In figure~\ref{fig:diffgc}, non-linear effects are responsible for the further decrease of the green curve for $k \geq 0.1\, h$/Mpc.

\item {\it Neutrino mass effects with $h$ varied to keep the CMB angular scales fixed: why does LSS data lifts the ($M_\nu$, $h$) degeneracy?}

The second part of the discussion consists in increasing $M_\nu$ by the same amount, while varying $h$ like in section~\ref{sec:cmb}, in such way as to keep a constant angular diameter distance to recombination, constant sound horizon angular scale, and constant damping angular scale (red dashed line). As we have seen in Section~\ref{sec:cmb} this procedure leads to the well known $(M_\nu,h)$ CMB degeneracy. 

We showed that this degeneracy is broken by BAO data, because the lower value of $h$ increases the angular diameter distance at low redshift (see Section~\ref{sec:bao}). This conclusion is valid also for galaxy $P(k)$ and shear $C_\ell^{ij}$, since the red dashed residuals in figures~\ref{fig:diffll}, \ref{fig:diffgc} are well outside the observational and theoretical error bars. For clarity, we should explain the shape of these red dashed lines, which is slightly counter-intuitive.

In the case of galaxy clustering, the higher value of $M_\nu$ and lower value of $h$ lead to an almost constant suppression of power on every scale, plus some wiggles on small scales (see figure~\ref{fig:diffgc}).
This may sound surprising since we are used to seeing more suppression on small scales when increasing the neutrino mass. This is true for fixed $h$, but here we are decreasing the Hubble rate at the same time. Since $\omega_\mathrm{m}=\Omega_\mathrm{m} h^2$ is kept fixed, this means that we are increasing $\Omega_\mathrm{m}$. For subtle reasons which can be understood analytically, the large-scale branch of the matter power spectrum is suppressed by the increase of $\Omega_\mathrm{m}$\footnote{In order to understand the observed behaviour, we have to elaborate on the matter power spectrum $P_{\rm m}$ entering equation~\ref{eq:Pg}. An analytic study of the linear power spectrum expressed in units of (Mpc$/h)^3$ as a function of $k$ in units of $h$/Mpc shows that at any given redshift, the large-scale branch ($k \ll k_\mathrm{eq}$) depends only on a factor $\left(g(\Omega_\mathrm{m},z)/\Omega_\mathrm{m}\right)^2$ coming from the Poisson equation and from the behaviour of matter perturbations during $\Lambda$ domination (see e.g. \cite{Lesgourgues:1519137}, equation (6.39)). The function $g(\Omega_\mathrm{m},z)\leq 1$ is related to the decrease of matter perturbations during $\Lambda$ domination. When increasing $\Omega_\mathrm{m}$, we decrease this factor $\left(g(\Omega_\mathrm{m},z)/\Omega_\mathrm{m}\right)^2$ and we suppress the large-scale power spectrum, but not the small-scale one. Indeed, looking again at equation (6.39) in \cite{Lesgourgues:1519137}, the small-scale branch receives an extra factor $\tilde{k}_{\rm eq}^4$ (i.e. $k_{\rm eq}^4$ with $k_{\rm eq}$ in $h$/Mpc). This new factor is actually proportional to $z_{\rm eq}^2 \Omega_{\rm m}^2$ (eq. (6.32) in the same reference), and the latter cancels the former $\Omega_{\rm m}^{-2}$ factor.}, while the small-scale branch is suppressed by massive neutrino free-streaming, coincidentally by roughly the same amount. This explains the almost constant suppression of power in the galaxy clustering spectrum (red dashed line, figure~\ref{fig:diffgc}). The wiggles located around $k\sim0.1\,h\mathrm{Mpc}^{-1}$ are related to the shift of the BAO scale due to the different angular diameter distance at low redshift, as we have explained in section~\ref{sec:bao} (see also reference~\cite{Poulin:2016nat}).

The situation is a bit different for the galaxy lensing spectrum $C_\ell^{ij}$ (red dashed line, figure~\ref{fig:diffll}) which probes metric fluctuations instead of matter fluctuations. As a result\footnote{Since the lensing spectrum directly depends on metric fluctuations,
it does not share with the matter power spectrum the factor $\Omega_\mathrm{m}^{-2}$ coming from the Poisson equation. Indeed, the factor $\Omega_\mathrm{m}^{-2}$ discussed in the previous footnote is exactly cancelled by a factor $\Omega_\mathrm{m}^2$ that appears in equation~\ref{eq:Cl} when replacing the window functions with equation~\ref{eq:W}. As a result, the large-scale branch of the $C_l^{ij}$'s depend on $g(\Omega_\mathrm{m},z)^2$ only, while the small-scale branch is proportional to $\Omega_\mathrm{m}^2$.}, the large-scale branch of the $C_l^{ij}$'s slightly increases when we decrease $h$ and increase $\Omega_\mathrm{m}$. Instead, the small-scale branch remains nearly constant due to the antagonist effects of neutrino free-streaming and of the increase in $\Omega_\mathrm{m}$, but the neutrino effect wins on non-linear scales. As can be seen on the right panel of figure~\ref{fig:diffll}, for the highest redshift bins, the lensing data is able to discriminate this effect and to lift the ($M_\nu$, $h$) degeneracy, although with less significance than the galaxy clustering data.

These conclusions are confirmed by the ($M_\nu$, $h$) joint probability contours presented in the upper right panel of figure~\ref{fig:euclidonly}, for CORE, Euclid-lensing and Euclid-pk. Indeed, the slope of the ($M_\nu$, $h$) degeneracy is different from the one observed in CMB data, and it is mainly driven by the CMB prior on $\theta_s$.

\item {\it Degeneracy between $M_\nu$ and other parameters from Large Scale Structure data alone.} 

Finally we increase $M_\nu$, decrease $h$ by a smaller amount than the one required for fixing $\theta_s$, and, at the same time, we vary the primordial power spectrum parameters, the amplitude $A_s$ and also the index $n_s$ in the case of cosmic shear (blue dotted lines). It is clear from figures~\ref{fig:diffll} and \ref{fig:diffgc} that this procedure can almost cancel the effect induced by a larger $M_\nu$ both in the shear $C_\ell^{ij}$ and in the galaxy $P(k)$, leading to a new degeneracy. We shall now explain the reasons for this degeneracy.

Considering that the primordial power spectrum of scalar perturbations is given by
\begin{equation}
\frac{k^3 \mathcal{P}_\mathcal{R}(k)}{2 \pi^2} =  A_s \left(\frac{k}{k_0}\right)^{n_s-1},
\label{eq:PR}
\end{equation}
the matter power spectrum $P_{\rm m}$ can be written as
\begin{equation}
P_{\rm m}(k,z) \propto A_s \left(\frac{k}{k_0}\right)^{n_s} T(k,z)^2
\label{eq:Pm}
\end{equation}
where $T(k,z)$ is the time and scale dependent linear transfer function of matter density fluctuations (not separable in the case of massive neutrinos).
\begin{figure*}[h]
\begin{tabular}{cc}
\includegraphics*[width=0.5\linewidth]{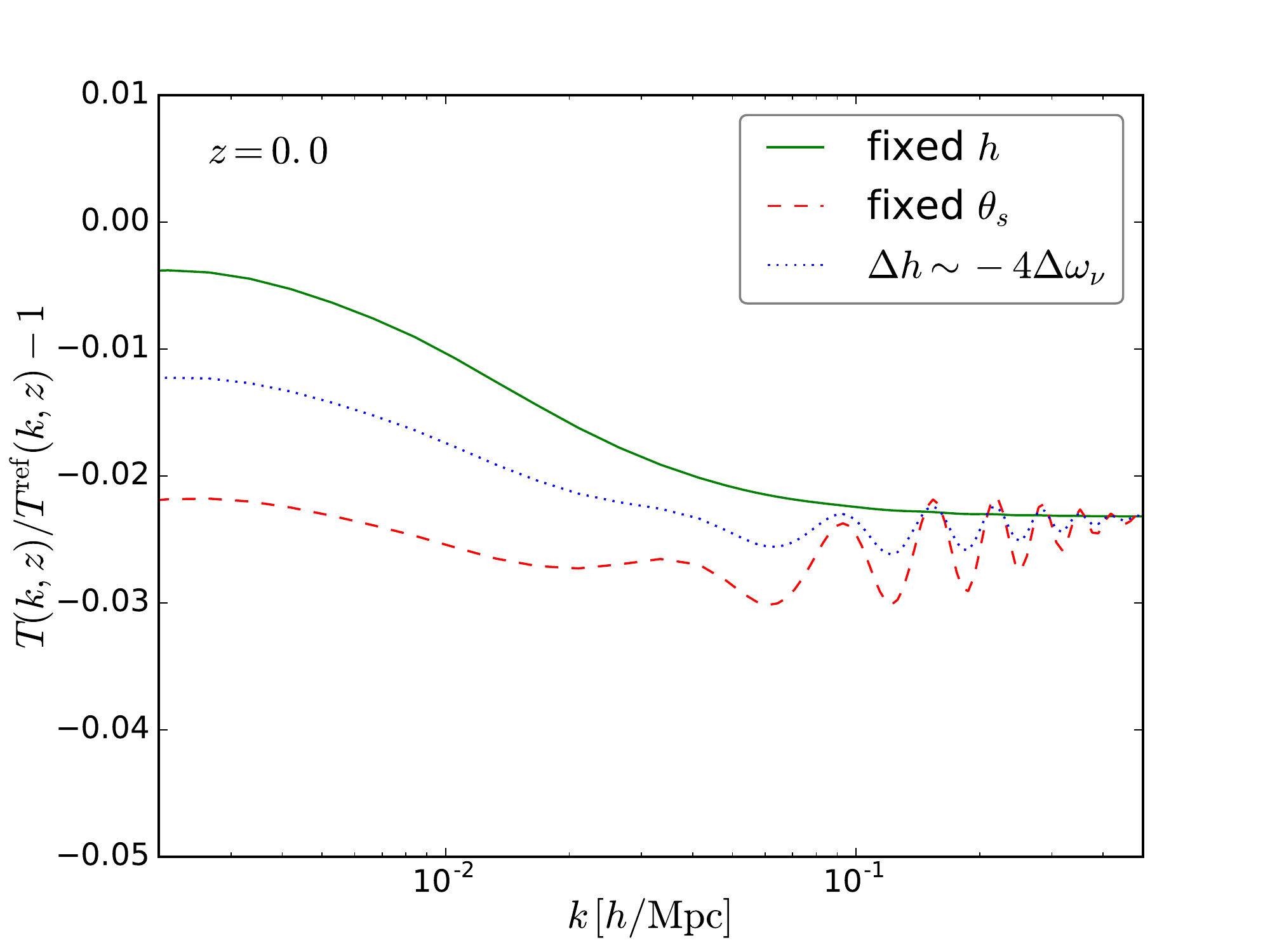} &
\includegraphics*[width=0.5\linewidth]{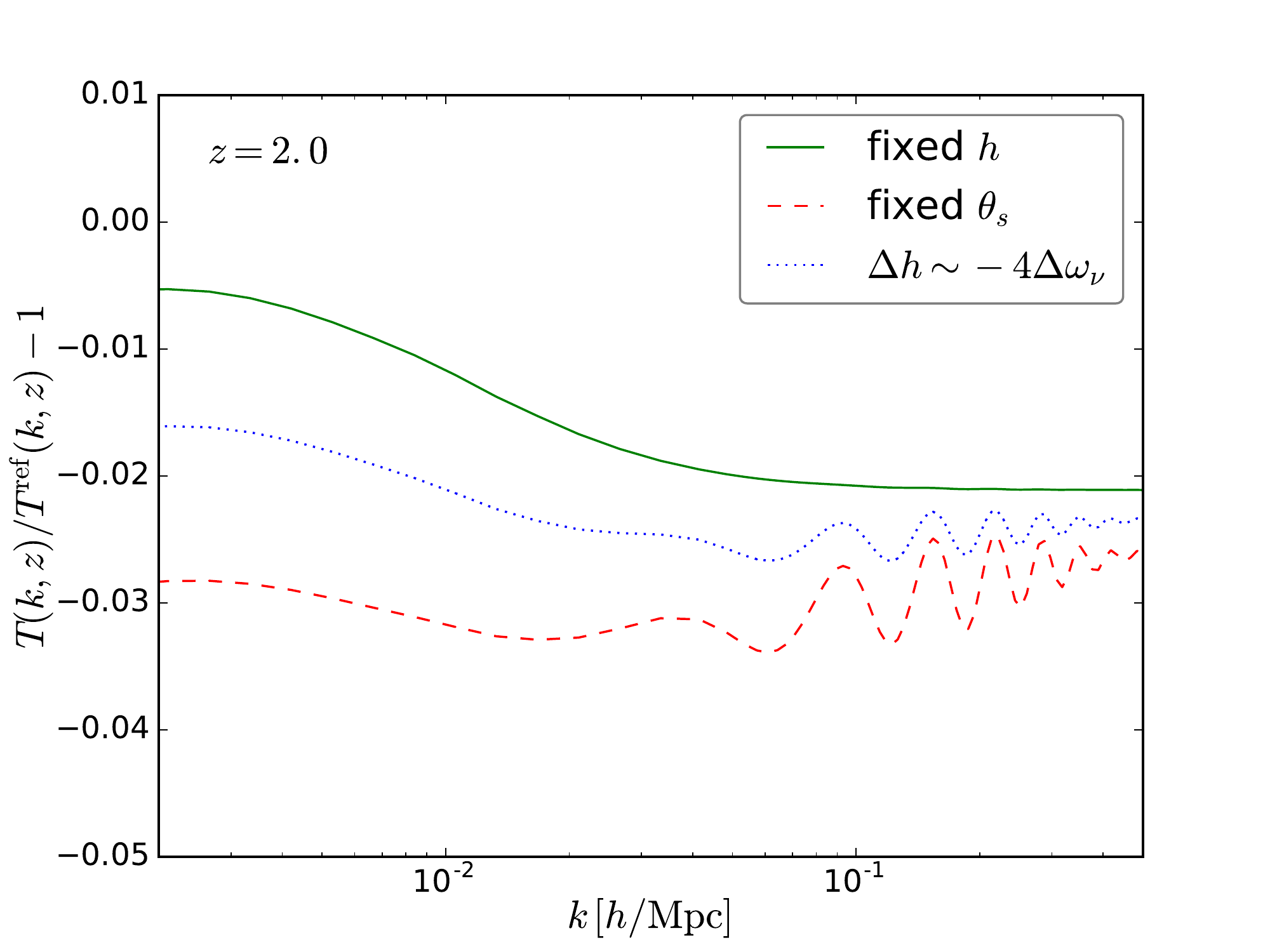} \\
\end{tabular}
\caption{ \label{fig:difftk}
Relative error on the linear transfer function $T(k)$ at redshift $z=0$ (left panel) and $z=2$ (right panel).
The line color/style - model correspondence is the same as in figure~\ref{fig:diffgc}.
}
\end{figure*}
As we have already explained, neutrinos induce a relative suppression of power on scales $k>k_\mathrm{fs}$; this suppression is encoded in the transfer function $T(k,z)$ of equation~\ref{eq:Pm}.
In figure~\ref{fig:difftk} we show how $T(k,z)$ is suppressed by a larger neutrino mass sum on $k>k_\mathrm{fs}$ at redshift $z=0$ and $z=2$. 
Changing $\left( n_s, A_s\right)$ affects only the primordial power spectrum, while leaving $T(k,z)$ unchanged, therefore, since we keep $\omega_b$ and $\omega_\mathrm{cdm}$ fixed, any deviation from the green solid line is due only to the variation of $h$ and $\Omega_m$. 
If, besides increasing $M_\nu$, we decrease $h$ to keep $\theta_s$ fixed (red dashed line), then the suppression of $T(k,z)$ extends to $k<k_\mathrm{fs}$ (because of the $\left(g(\Omega_\mathrm{m},z)/\Omega_\mathrm{m}\right)^2$ factor) and the wiggles, due to the shift of the BAO scale, appear at smaller scales. This graphically explains what we have already discussed in point 2.
Reducing the tweaking on $h$ (blue dotted line) implies less reduction of power on the large scale branch and a smoothing of the wiggles; anyhow, the massive neutrino suppression of the transfer function is not fully compensated. 
However, if we look at equation~\ref{eq:Pm} it is clear that a red tilt of the primordial power spectrum, combined with a smaller normalization, can mimic the same effect of a larger neutrino mass, reducing power on small scales respect to large scales.

The left and right bottom panels of figure~\ref{fig:euclidonly} show the degeneracies between $M_\nu$ and $(n_s,A_s)$. We can see that the degeneracy between $M_\nu$ and $n_s$ is mildly positive in galaxy lensing, as expected from the discussion above, while it is negative in CMB, as explained at the end of section~\ref{sec:cmb}, and mildly negative in galaxy clustering.
The reason why this positive $(M_\nu,n_s)$ correlation emerges with cosmic shear, but not with galaxy correlation data, is related to the window function.
Indeed, since the window function (equation~\ref{eq:W}) for each redshift bin is given by the integral over the line of sight, the $C_\ell^{ij}$'s of equation~\ref{eq:Cl} receive contributions from a larger range of scales. Therefore,
being sensitive to a wider lever arm in $k$ space, cosmic shear will be particularly sensitive to scale dependent variations of the power spectrum.

Notice that here the tweaking of $A_s$ is larger than the one we performed at point 3 of section~\ref{sec:cmb}. Thus, the corresponding $\Delta \tau_\mathrm{reio} \sim 0.5 \ln(1.05) \sim 0.027$ would lead to an enhancement of the reionization bump even bigger than the one we observed in the blue dotted line of the $C_\ell^{EE}$ plot (figure~\ref{fig:difftt}, second row, right panel). This already shows that the degeneracy discussed here can be lifted by combining LSS data with CMB data. Nevertheless this discussion was important to understand the pulls in parameter space appearing when all data sets are combined with each other.

\end{enumerate}

Figure~\ref{fig:euclidonly} confirms the points discussed previously, and provides a comprehensive graphical summary of  the complementarity between future CMB and LSS data in the context of neutrino mass measurement. 

First, we see that even when adopting CMB-derived priors on $\omega_b$ and $\theta_s$, the LSS data cannot efficiently constrain the neutrino mass, due to the degeneracy discussed in the previous paragraphs (point 3), involving mainly ($M_\nu$, $A_s$, $H_0$), and to a lesser extent, $n_s$. We have seen that this degeneracy requires a milder correlation between $M_\nu$ and $H_0$ than the CMB data: $\Delta h \sim - 3 \Delta \omega_\nu$ for LSS alone, instead of $\Delta h \sim - 12 \Delta \omega_\nu$ for CMB alone. Since in Figure~\ref{fig:euclidonly} the Euclid mock data was fitted together with a prior on $\theta_s$, the final correlation angles represent compromises between these values. The lensing data also exhibits a negative correlation between $M_\nu$ and $\omega_{\rm cdm}$.

The CMB and LSS contours of Figure~\ref{fig:euclidonly} clearly intersect each other for several pairs of parameters:
\begin{itemize}
\item The CMB and LSS data prefer different directions of degeneracy in ($M_\nu$, $H_0$) space, hence the combination between them can strongly reduce the uncertainty on both $M_\nu$ and $H_0$.
\item 
The CMB data lifts the ($M_\nu, A_s$) degeneracy present in the LSS data, for the reason mentioned above: the shift in $A_s$ would need to be compensated by a shift in $\tau_{\rm reio}$ producing a reionisation bump incompatible with the data. However, in the combined data set, the LSS data would keep pulling towards more positive correlation between  $M_\nu$ and $A_s$.
\item
the very different correlations in ($M_\nu$, $\omega_{\rm cdm}$) space reduces uncertainties on $\omega_{\rm cdm}$, with a side effect on the CMB side. We have seen that the effect of neutrino masses on the CMB lensing spectrum can be compensated either by playing with $\omega_{\rm cdm}$, or with ($A_s, \tau_{\rm reio}$). The CMB alone would favour the first option. Like BAO data, weak lensing data breaks the ($M_\nu$, $\omega_{\rm cdm}$) degeneracy and leaves only the second option. This goes in the same direction as the previous point: pulling towards more positive correlation between  $M_\nu$ and $A_s$.
\end{itemize}
Hence we can already anticipate that the combination of CMB plus LSS data leads to an enhanced degeneracy between ($M_\nu, A_s$) compared to CMB data alone. As a consequence, in order to maintain a fixed combination $A_s e^{-2 \tau_{\rm reio}}$, the combined data may generate a significant correlation in ($M_\nu, \tau_{\rm reio}$) space. 

The goal of the next section is to check these partial conclusions with a global fit of all data sets at the same time.

\section{Joint Analysis Results}
\label{sec:results}

\subsection{Combination of CMB, BAO and galaxy shear/correlation data}
\label{sec:results-without-21}

In this section we will present the results of our Markov Chain Monte Carlo forecast of the combined sensitivity of future CMB, BAO and LSS experiments to the cosmological parameters described in Section~\ref{sec:cmb_intro}, in particular to the neutrino mass sum. As already mentioned in Section~\ref{sec:cmb}, our MCMC forecast will be performed using the {\sc MontePython} code\footnote{\tt http://baudren.github.io/montepython.html}~\cite{Audren:2012wb}, interfaced with the Boltzmann solver {\sc class}\footnote{\tt http://class-code.net}~\cite{Lesgourgues:2011re,Blas:2011rf,Lesgourgues:2011rh}. 
We already commented at the end of section~\ref{sec:lss-spec} our conservative choices for the precision parameters:  theoretical error parameter $\epsilon=0.025$, cut-off at  $k_\mathrm{max}=0.6\, h/$Mpc for galaxy correlation, and at $\ell_\mathrm{max}=2000$ for cosmic shear. Still this choice comes from a subjective estimate of the accuracy with which non linear corrections and systematic effects will be modelled in the future, and different assumptions would lead to different parameter sensitivities.

\begin{table}[th]
\resizebox{1.0\textwidth}{!}{%
\begin{tabular}{| l | c | c | c | c | c | c |}
\hline
&$\sigma(M_\nu)/[\mathrm{meV}]$&$\sigma(\tau_{\rm reio})$&$\sigma(10^9A_s)$&$\sigma(n_s)$&$\sigma(\omega_{\rm cdm})$&$\sigma(h)$ \\
\hline
CORE&$42$&$0.0020$&$0.0084$&$0.0018$&$0.00052$&$0.0052$\\
CORE+DESI&$19$&$0.0020$&$0.0080$&$0.0014$&$0.00026$&$0.0022$\\
CORE+DESI+Euclid-lensing&$16$&$0.0020$&$0.0078$&$0.0014$&$0.00023$&$0.0019$\\
CORE+Euclid (lensing+pk)&$14$&$0.0020$&$0.0079$&$0.0015$&$0.00025$&$0.0017$\\
CORE+Euclid (lensing+pk)+21cm&$12$&$--$&$0.0042$&$0.0014$&$0.00021$&$0.0017$\\
\hline
\end{tabular}%
}
\caption{Expected $1\,\sigma$ sensitivity of CORE, CORE + DESI, CORE + DESI + Euclid (lensing), CORE + Euclid (lensing+pk), 
CORE + Euclid (lensing+pk) + ``21cm-motivated $\tau_{\rm reio}$ prior''
to the parameters $\{M_\nu, \tau_{\rm reio}, 10^9A_s, n_s, \omega_{\rm cdm}, h\}$.
We did not combine DESI and Euclid-pk in order to avoid double counting the information coming from the wiggly part of the spectrum.
}
\label{tab:errors}
\end{table}

\begin{figure*}[h]
\centering
\includegraphics*[width=0.9\linewidth]{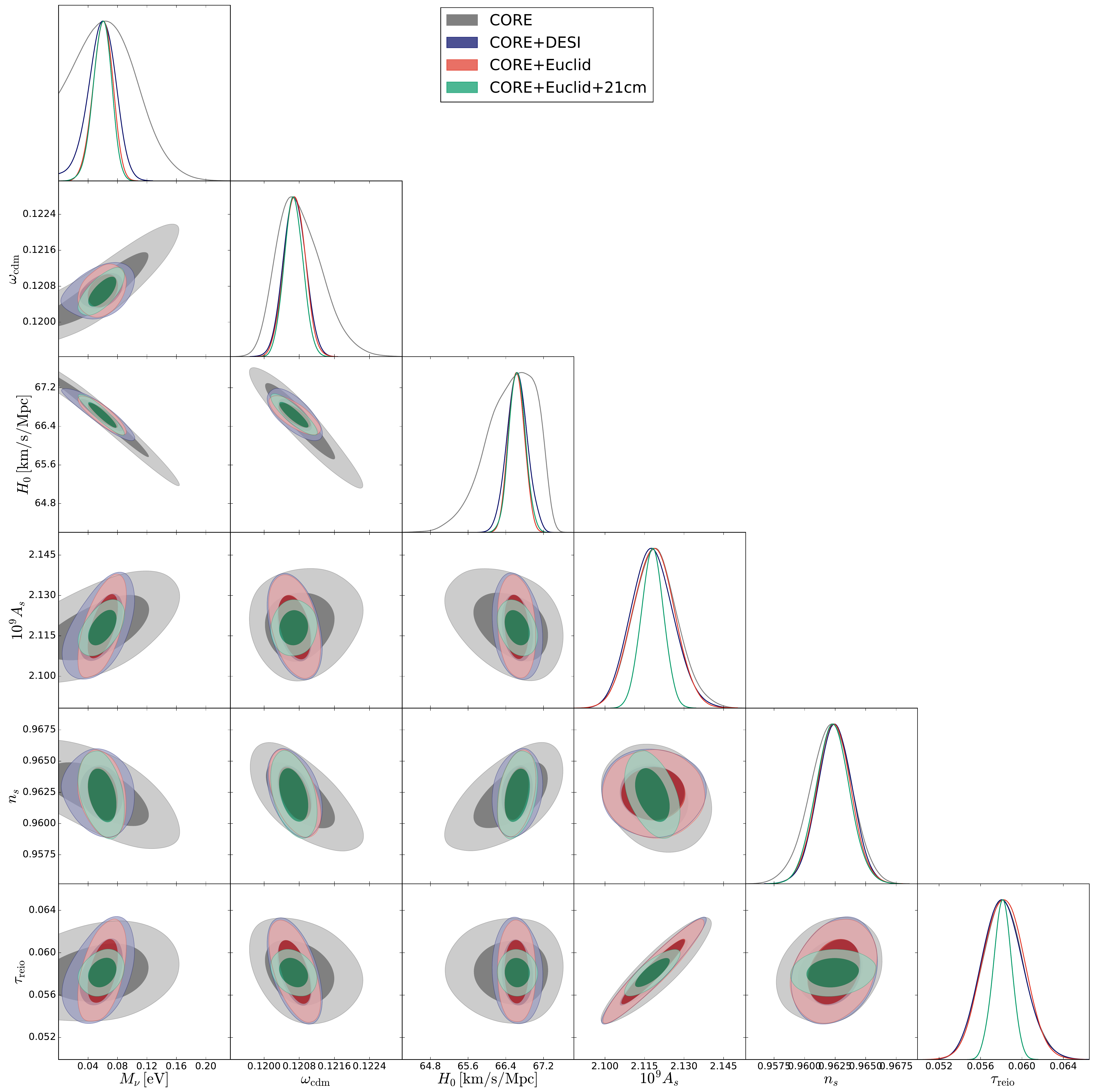}
\caption{ \label{fig:triangle}
Marginalized one$-$ or two$-\sigma$ contours and one dimensional posteriors in the $\left(M_\nu, \omega_\mathrm{cdm}, H_0, A_s, n_s, \tau_{\rm reio}\right)$ parameter space, showing the expected sensitivity of various future experiments: CORE only (gray contours), CORE+DESI (blue contours), CORE+Euclid (red contours) and CORE+Euclid+21cm (green contours).
The last independent parameter, $\omega_\mathrm{b}$, is always very well constrained by CMB data alone.
}
\end{figure*}

In the first four lines of table~\ref{tab:errors} we report the expected sensitivity of CORE, CORE+DESI, CORE+DESI+Euclid-lensing and CORE+Euclid (lensing+pk)\footnote{Contrarily to an earlier version of this work, to avoid any possible ``double counting'' of the BAO information, we will not combine DESI and Euclid-pk data.} to $M_\nu$ and other cosmological parameters playing a crucial role in our analysis of parameter degeneracies: $\tau_{\rm reio}$, $10^9A_s$, $n_s$, $\omega_{\rm cdm}$ and $h$ (the last independent parameter, $\omega_\mathrm{b}$, is always very well constrained by CMB data alone). In figure~\ref{fig:triangle} we plot the one dimensional posteriors and the one- and two-$\sigma$ marginalized contours for the same parameters.

First of all we notice that the projected $1\,\sigma$ errors in table~\ref{tab:errors} and 1D distributions in figure~\ref{fig:triangle} reflect the theoretical points we have discussed in the previous sections: both DESI and Euclid greatly improve the sensitivity to $M_\nu$, $\omega_\mathrm{cdm}$ and $h$. The uncertainty on $M_\nu$ tightens by more than a factor two for CORE+DESI and a factor three for CORE+Euclid, compare to the CORE only sensitivity. 
The error on $H_0$ shrinks by a factor larger than two for CORE+DESI and a factor three for CORE+Euclid.
However once more we want to stress that in the case of DESI the improved sensitivity arises from reducing the degeneracy between $H_0$ and $M_\nu$, while in the case of Euclid the longer lever arm of the shear data is specifically sensitive to the suppression of power at small scales induced by $M_\nu$.

The first column of figure~\ref{fig:triangle} shows all the degeneracies with respect to $M_\nu$. Let us describe the evolution of those correlations with the addition of the different datasets:
\begin{enumerate}
\item {\it CORE data only}. When only CMB data are considered, correlations follow the directions expected from our extensive discussion of section \ref{sec:cmb}. Let us just note that contrarily to $\Lambda$CDM runs {\em without} neutrino mass as a free parameter, the mild correlation between $A_s$ and $n_s$ is negative, which is a result of the mild negative (resp. positive) correlation between $M_\nu$ and $n_s$ (resp. $A_s$).

\item {\it Adding DESI data.}  In general, the size of the 2D-distributions shrink by a factor $\sim$2.
The extended regions defining the positive correlations between ($M_\nu, A_s$) and ($M_\nu, \tau_{\rm reio}$) become steeper, since it is not possible anymore to play with $H_0$ or $\omega_{\rm cdm}$ to compensate the effect of the summed neutrino mass on the CMB lensing spectrum.
Indeed, as described in section~\ref{sec:bao}, moving along this degeneracy direction would lead to very different BAO angular scales. Thus, the effect of the summed neutrino mass on CMB lensing is rather compensated by playing with parameters to which BAO data are insensitive\footnote{As side remarks, note that such compensation cannot be done by playing with $n_s$: as a consequence, both the ($M_\nu,\,n_s$) degeneracy and the ($A_s, \,n_s$) degeneracy are lifted when BAO data are added;
finally, because of the different neutrino mass compensation driven by the inclusion of BAO data, the correlations of $\omega_\mathrm{cdm}$ and $H_0$ with respect to $A_s,n_s,\tau_\mathrm{reio}$ are lifted, as well.}, namely $A_s$ and $\tau_{\rm reio}$. 

\item {\it Adding Euclid (lensing + $P(k)$) data}. Most of the discussion on the inclusion of DESI data still applies here, since Euclid data contains information on the BAO scale at different redshift. However the matter / shear power spectra contain extra information on cosmological perturbations, and lift or reinforce some parameter degeneracies, consistently with our previous discussion in section~4.2, point 3. The ($M_\nu, H_0$) degeneracies get considerably reduced because the LSS data would prefer a different correlation angle between these two parameters. As expected, the Euclid data considerably tightens the positive correlation between $M_\nu$ and $A_s$, and as a side effect the combined data leads to a clear positive correlation between $M_\nu$ and $\tau_{\rm reio}$. The degeneracy between $M_\nu$ and $\omega_{\rm cdm}$ is lifted by the weak lensing data. All these degeneracy reductions lead to an overall shrinking of all contours involving $M_\nu$, $H_0$ and $\omega_{\rm cdm}$ by a factor of order 3 between CMB and CMB+LSS data. The neutrino mass value is accurately determined independently of the value of $n_s$, and the mild correlation between $M_\nu$ and $n_s$ in CMB data disappears with additional LSS data.
\end{enumerate}

Compared to figure~\ref{fig:triangle}, table~\ref{tab:errors} presents the results of one more MCMC run featuring CORE, DESI and Euclid weak lensing, but not the Euclid galaxy clustering information. The comparison of these results with those for CORE and Euclid weak lensing+galaxy clustering show the importance of geometrical information (BAO angular scales) versus shape information (full matter power spectrum), although both runs do contain some shape information coming from the weak lensing data. We clearly see that adding more shape information on the matter power spectrum benefits only to the determination of $M_\nu$ and $H_0$, and actually by a modest amount (10 to 15\% per cent). At face value, this means that even if the analysis of future galaxy clustering data was plagued by unexpected systematics (besides the level that we conservatively took into account with our theoretical error bar), the prospects to accurately determine the summed neutrino mass with future surveys would not collapse.

In order to further improve the measurement of the neutrino mass with cosmological data, one should try to add independent constraints on the parameters that remain most strongly correlated with $M_\nu$ in the CMB+LSS contours: these are $H_0$, $A_s$ and $\tau_\mathrm{reio}$. The role of a very precise determination of $H_0$, free of astrophysical systematics, for the measurement of the neutrino mass, has already been stressed e.g. in~\cite{Moresco:2016nqq}.
It was also previously noticed in Ref.~\cite{Liu:2015txa} that 21cm surveys could improve the determination of the optical depth to reionization, and thus of the summed neutrino mass. Having understood the physical explanation for the ($M_\nu, \tau_{\rm reio}$) degeneracy, we wish to further investigate this possibility, while keeping our conservative assumption on the matter power spectrum theoretical error.

\subsection{Adding 21cm surveys}


In the near future, many experimental efforts will be devoted to measuring precisely the epoch of recombination (EoR), mostly through the 21 cm line created by the hyperfine transition of the Hydrogen atom\footnote{e.g. PAPER 64: \url{http://eor.berkeley.edu}, 21CMA: \url{http://21cma.bao.ac.cn}, MWA: \url{http://www.mwatelescope.org}, LOFAR: \url{http://www.lofar.org}, HERA: \url{http://reionization.org} or SKA: \url{http://www.skatelescope.org.}}, including the value of $\tau_{\rm reio}$. In general,  details of the EoR are strongly connected to fundamental questions in cosmology and astrophysics. They could shed light on many properties of the first galaxies and quasars, measure the time at which they form, explain how the formation of very metal-poor stars proceeded,  and reveal whether the first galaxies were indeed the only re-ionizing source.

However, these experiments can also have great implications for neutrino physics in cosmology. Indeed, the independent measurement of the epoch of reionization by 21cm surveys may break the degeneracy between $A_s$ and $\tau_{\rm reio}$~\cite{Villaescusa-Navarro:2015cca,Liu:2015txa} which appears in combined analyses of future CMB+LSS data. 

To assess the impact of 21cm surveys on $\sigma(M_\nu)$,
we performed a final MCMC run combining CORE+Euclid mock data with a gaussian prior on the value of $\tau_{\rm reio}$.
In agreement with forecasts on the sensitivity of HERA or SKA, we fixed the prior variance to $\sigma(\tau_{\rm reio}) = 0.001$~\cite{Santos:2015gra,Liu:2015txa}. Note that by doing so, we are being conservative, since 21cm surveys will not only measure the evolution of the mean free electron fraction $x_e(z)$  (and thus the optical depth $\tau_{\rm reio}$), but also the power spectrum  of the 21cm signal at different redshifts, $P_{21cm}(k,z)$, related to variations along the line of sight of the free electron fraction $x_e(\hat{n},z)$ \cite{Mellema:2012ht}. We are therefore using the minimal amount of information that one can extract from these experiments, and one could go beyond following e.g. the procedure of Refs.~\cite{Oyama:2012tq,Liu:2015txa,Villaescusa-Navarro:2015cca,Oyama:2015gma}. 

The results of our MCMC forecast are summarized by the last line of table~\ref{tab:errors}, and the green contours in figure~\ref{fig:triangle}.

The main impact of the $\tau_{\rm reio}-$prior is to reduce the possibility of varying of $A_s$, necessary to adjust the CMB parameter $A_s\exp(-2\tau_{\rm reio})$, by almost a factor two.
Since $M_\nu$ was correlated directly with $A_s$ and indirectly with $\tau_{\rm reio}$, the sensitivity to the summed neutrino mass also improves thanks to 21cm data, going from $\sigma(M_\nu) = 14$~meV for CORE+Euclid to $12$~meV. As a side effect, the positive correlation between $M_\nu$ and $\omega_\mathrm{cdm}$ and the negative correlation between $M_\nu$ and $h$ get steeper.


Thus, even if nature has chosen the summed neutrino mass to be close to the lower limit of the normal hierarchy, $M_\nu=60$~meV, we expect that the joint analysis of CORE + Euclid + 21cm data will detect it at more than $5~\sigma$.
\section{Conclusions}
\label{sec:conclusions}

The foundations of a new era in precision cosmology are based on two cornerstones: the high sensitivity of future CMB and galaxy survey experiments, and a deep understanding of the physics governing the processes of recombination and structure formation. The extreme accuracy of future data will offer the opportunity to constrain particle physics with cosmology, exceeding in many cases the precision of laboratory experiments. However, in order to exploit the new data, cosmologists will need an accurate enough theoretical model taking into account the underlying physics.

Neutrinos provide an excellent example of how the sensitivity of future cosmological surveys may lead to such an important result as the summed neutrino mass detection, even when uncertainties on the details of the cosmological model are marginalised over.

In this paper we have provided a careful discussion of the physical effects induced by massive neutrinos and their impact on cosmological observables, as they will appear in the data analysis of the next generation of cosmological experiments. We have shown how the unique nature of light neutrinos, being relativistic until very late times and behaving as a matter component after the non relativistic transition, makes possible to identify different signatures at different epochs of the cosmic history. Therefore the correlation between the summed neutrino mass and the other cosmological parameters changes, depending on the redshift range probed by the various data sets.  

Our results on the sensitivity of future CMB-CORE and BAO-DESI experiments to the summed neutrino mass are consistent with the literature (see Refs.~\cite{Allison:2015qca, Liu:2015txa}). Moreover, the results of our forecasts including a Euclid-like survey prove the importance of cosmic shear and galaxy clustering as complementary probes. 
We pointed out that the results of our Euclid cosmic shear + galaxy correlation forecasts depend very much on the choice of the theoretical error introduced to account for the systematics coming from the deep non-linear regime. Nevertheless, they are again compatible with previously published results. For instance, Ref.~\cite{Hamann:2012fe} found 
$\sigma(M_\nu)=11$~meV for Planck + Euclid cosmic shear / galaxy correlation, but with a different treatment of the uncertainty on non linear corrections. Ref.~\cite{Audren:2012wb} found a larger error, close to 20~meV, but for Planck + Euclid cosmic shear or Planck + Euclid galaxy correlation, not trying to combine the two LSS probes together and without CORE data. Ref.~\cite{Liu:2015txa} found $\sigma(M_\nu)=12$~meV for Planck + CMB-Stage-IV + BAO-DESI + 21cm-HERA, identical to our estimate for CORE + Euclid + 21cm-$\tau_{\rm reio}$-prior.

Anyhow, the main goal of this study was not to present a new set of forecasts, but to discuss the details of physical effects and parameter degeneracies involving neutrino masses. In particular, we clarified the reason for which 
an unexpected degeneracy between the neutrino mass sum and the optical depth at reionization will appear in the analysis of future high precision galaxy surveys, as already pointed out e.g. in \cite{Liu:2015txa,Allison:2015qca}. We showed that this degeneracy is not present in a CMB-only analysis, because the neutrino mass effects on CMB lensing can be compensated by playing with $h$ and $\omega_\mathrm{cdm}$ in a better way than by adjusting $(A_s, \tau_\mathrm{reio})$. However, the former degeneracy is lifted once BAO and LSS low redshift measurements are taken into account. Moreover, we demonstrated that the LSS data introduce a strong correlation between $M_\nu$ and $A_s$, which finally leads to a clear ($M_\nu, \tau_\mathrm{reio}$) degeneracy in the combined CMB+LSS analysis.

These conclusions clarify why further independent measurements of  the optical depth will benefit to the neutrino mass determination, as previously noticed by the authors of~\cite{Liu:2015txa}.
For instance, the results from the HERA or SKA 21cm surveys will provide an independent constraint on $\tau_\mathrm{reio}$, thus breaking this degeneracy. Our results indicate that this could reduce the error on $M_\nu$ with respect to the CMB+LSS case, leading to a robust detection of the summed neutrino mass at more than $5~\sigma$ for CORE+Euclid+HERA or SKA. In principle, it would be possible to do even better if $H_0$ could be measured in an independent and robust way with an error below $\sigma(H_0) \sim 0.17$km/s/Mpc.

In conclusion, the remarkable complementarity of future different cosmological data will lead to extremely accurate constraints on the neutrino mass sum and, possibly, on other neutrino properties, answering some of the still open questions of modern physics.

\section*{Acknowledgement}
JL acknowledges extremely useful discussions with A. Challinor.
VP is supported by the ``Investissements d'avenir, Labex ENIGMASS'',  of  the  French  ANR.
This work has been done thanks to the facilities offered by the RWTH High Performance Computing cluster, and the Universit\'e Savoie Mont Blanc MUST computing center.

\bibliography{biblio}{}
\bibliographystyle{plain}

\end{document}